\documentclass[prb,aps,twocolumn,showpacs]{revtex4}
 \usepackage{graphicx}

\begin{document}

\title{Scanning SQUID susceptometry of a paramagnetic superconductor}

\author{J.R. Kirtley$^1$, B. Kalisky$^1$, J.A. Bert$^2$, C. Bell$^2$, Y. Hikita$^2$, H.Y. Hwang$^2$, J.H. Ngai$^3$, Y. Segal$^3$, F.J. Walker$^3$, C.H. Ahn$^3$ and K.A.  Moler$^{1,2}$}

\affiliation{$^1$Departments of Physics and Applied Physics, Stanford University, Stanford, CA 94305, USA}
\affiliation{$^2$Stanford Institute for Materials and Energy Science, Stanford University, Stanford, California 94305, USA}
\affiliation{$^3$Departments of Applied Physics and Physics, Yale University, New Haven, CT 06520, USA}
\begin{abstract}
Scanning SQUID susceptometry images the local magnetization and susceptibility of a sample. By accurately modeling the SQUID signal we can determine physical properties such as the penetration depth and permeability of superconducting samples.  We calculate the scanning SQUID susceptometry signal for a superconducting slab of arbitrary thickness with isotropic London penetration depth $\lambda$, on a non-superconducting substrate, where both slab and substrate can have a paramagnetic response that is linear in the applied field. We derive analytical approximations to our general expression in a number of limits. Using our results, we fit experimental susceptibility data as a function of the sample-sensor spacing for three samples: 1) $\delta$-doped SrTiO$_3$, which has a predominantly diamagnetic response, 2) a thin film of LaNiO$_3$, which has a predominantly paramagnetic response, and 3) the two-dimensional electron layer (2-DEL) at a SrTiO$_3$/LaAlO$_3$ interface, which exhibits both types of response. These formulas will allow the determination of the concentrations of paramagnetic spins and superconducting carriers from fits to scanning SQUID susceptibility measurements.
 \end{abstract}

\maketitle

\section{Introduction}


Scanning SQUID microscopy \cite{kirtley1999ssm,kirtley2010fundamental} allows the simultaneous imaging of the local magnetization and the magnetic response (susceptibility)\cite{gardner2001scanning} of the surface of a sample on a micron length scale. The sign and magnitude of the susceptibility signal yields information about electrons in the material. For a superconductor, the diamagnetic susceptibility is a measure of the local London penetration depth.\cite{tafuri2004magnetic,hicks2009evidence,luan2011local,kalisky2010stripes}  In most superconductors, the diamagnetic susceptibility is much stronger than other possible sources of magnetic response, such as nuclear susceptibility or the paramagnetism of impurities, other regions of the sample, or non-superconducting carriers.  However, in superconductors with unusually strong competing paramagnetic  susceptibility and/or a low superfluid density, it may be necessary to consider both types of contributions.  For example, a paramagnetic response has  been observed in scanning susceptometry measurements of non-superconducting samples\cite{bluhm2009spinlike} and superconducting samples above their critical temperatures.\cite{bert2011direct,kalisky2011para} 

The temperature dependence of the London penetration depth, which is related to the susceptibility, has played an important role in determining the symmetry of the superconducting order parameter in unconventional superconductors.\cite{prozorov2006magnetic}  However, for superconductors with low superfluid densities, the diamagnetic contributions from Cooper pairs and the paramagnetic contributions from spin or other sources can have similar magnitudes but different temperature dependences, making it difficult to determine the temperature dependence of the superfluid density.\cite{prozorov2000evidence} It is therefore important to be able to separate the paramagnetic from the superconducting components in scanning SQUID susceptometry measurements.

Kogan\cite{kogan2003meissner} presented a model for the diamagnetic response of a superconductor to arbitrary local field sources.  One source he considered was a circular ring of current appropriate for scanning SQUID susceptometry. Here we extend his model to include both diamagnetic and paramagnetic effects, for a paramagnetic superconductor of arbitrary thickness on a paramagnetic substrate. Our final expression reduces to that of Kogan\cite{kogan2003meissner} for a superconductor with the permeability of vacuum in the bulk and thin film limits, and to that of Bluhm {\it et al.} \cite{bluhm2009spinlike} for a thin film paramagnetic response. We present in Table \ref{tab:analytical_limits} analytical approximations to our full expression for a) a bulk non-superconducting paramagnet, b) a thin non-superconducting paramagnet, c) a bulk superconductor without paramagnetism with penetration depth short relative to the other lengths in the problem, d) a bulk superconductor without paramagnetism with penetration depth long relative to other lengths in the problem, and e) a thin superconductor without paramagnetism. These analytical approximations, along with the full expression Eq. \ref{eq:full_susc_expression}, represent the main results of this paper.

Although in this paper we concentrate on the scanning SQUID susceptometer geometry, with the field coil co-planar and co-axial with the pickup loop , the same basic formalism could be applied to penetration depth measurements using other two-coil mutual inductance geometries,\cite{fiory1988penetration,lee1994crossover,claassen1997optimizing} for example, with the two coils at different heights, or on opposite sides of a thin film sample.

As examples of applications of these expressions we fit scanning susceptometry data on a $\delta$-doped sample of SrTiO$_3$, a thin film of LaNiO$_3$, and the two-dimensional electron layer (2-DEL) at the interface between SrTiO$_3$ and LaAlO$_3$.\cite{ohtomo2004high} For the case of $\delta$-doped SrTiO$_3$, which has a predominantly diamagnetic response, there is a strong correlation between the best penetration depth and sensor height parameters in fitting susceptibility approach curves: an uncertainty in the sensor height of 1.5$\mu$m results in an uncertainty in the Pearl penetration depth $\Lambda=2\lambda^2/t$ of 60\%. A susceptibility approach curve for the LAO/STO interface at a position which is  predominantly paramagnetic is well fit by our expression for a thin film paramagnet with reasonable values for the fitting parameters. However, at a position which is weakly diamagnetic the best fits for the height and permeability parameters take on unphysical (negative) values, even for a model which includes both superconducting and paramagnetic contributions. We speculate that this last may be due to sample inhomogeneity and/or an interaction between the SQUID and the sample superfluid.


\begin{figure}
\includegraphics[width=3.3in,trim=0 0 0 0]{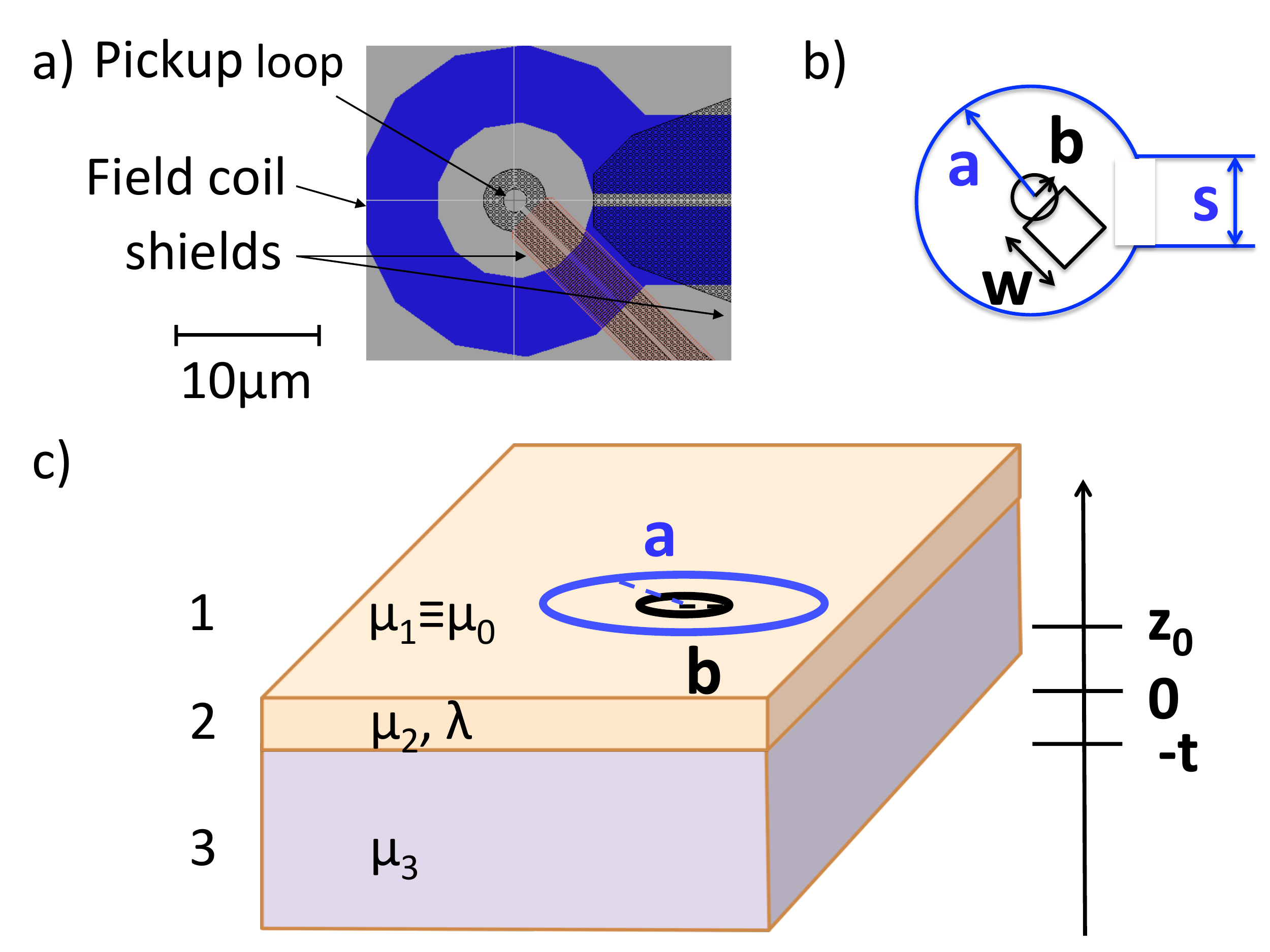}
\caption{Model geometry.  a) The layout of the field coil/pickup loop region of the SQUID susceptometers used in this paper. b) Approximations to this layout considered in Appendices \ref{sec:field_coil_approx} and \ref{sec:pickup_loop_approx}. c) For this paper we consider a slab of thickness $t$ with magnetic permeability $\mu_2$ and superconducting penetration depth $\lambda$ on a non-superconducting substrate with magnetic permeability $\mu_3$.   We define $z=0$ at the sample surface, with a pickup loop (radius b) and concentric field coil (radius a) located at a height $z = z_0$ in a plane parallel to the sample surface.     }.
\label{fig:slab_susc_geo}
\end{figure}

\section{Model}
\subsection{Full expression}

We consider the geometry of Figure \ref{fig:slab_susc_geo}. The SQUID susceptometers used in this paper have the layout shown in Fig. \ref{fig:slab_susc_geo}a.\cite{huber2008gradiometric} Traditionally such a layout has been approximated by that of Fig. \ref{fig:slab_susc_geo}b: the field coil is represented by a circular wire, while the pickup loop is represented by a circular wire plus an additional pickup area due to flux redirection from the leads.\cite{ketchen1995design} In this paper we assume the geometry of Fig. \ref{fig:slab_susc_geo}c: the susceptometer is represented by two co-planar concentric circular loops. The field coil has radius $a$, and the pickup loop has radius $b$ . Both are infinitely thin wires. We evaluate in Appendices \ref{sec:field_coil_approx} and \ref{sec:pickup_loop_approx} the systematic errors in the SQUID susceptibility associated with our approximations to the field coil and pickup loop shapes. The loops are oriented parallel to, and a height $z_0$ above, a slab of material with thickness $t$, permeability $\mu_2$ and isotropic London penetration depth $\lambda$ on a semi-infinite non-superconducting substrate with permeability $\mu_3$. 

One can divide space into 3 regions of interest: 1: $z_0>z>0$, 2: $0>z>-t$, and 3: $-t>z$. The half-space $-t>z$ has permeability $\mu_3$ and is non superconducting. The space $t>z>0$ has $\mu_1=\mu_0$ and is also not superconducting.   

Following Kogan,\cite{kogan2003meissner} $\nabla\times\vec{H}_1=0$ in region 1 since there are no currents. This identity is assured by writing the field $\vec{H}_1$ as the gradient of a scalar potential $\varphi_1$: $\vec{H}_1=\nabla\varphi_1$.  Then $\nabla^2\varphi_1=0$ since $\nabla \cdot \vec{B}_1=\nabla \cdot (\mu_0\vec{H}_1)=0$. In region 3 there is again no transport current, so $\nabla \times \vec{H}_3=0$. Since we consider only the case where $\mu$ is spatially homogeneous, $\nabla \cdot \vec{H}_3=0$, and therefore $\nabla^2\varphi_3=0$.

Inside the superconductor the total current, proportional to the curl of the magnetic flux density $\vec{B}$, consists of the supercurrent plus the current due to the inhomogeneity of the magnetization $\vec{M}$:
\begin{equation}
\nabla \times \vec{B} = \frac{\Phi_0}{2\pi\lambda^2} ( \nabla \theta + \frac{2\pi\vec{A}}{\Phi_0})+\mu_0 \nabla \times \vec{M},
\label{eq:total_current}
\end{equation}
where the London penetration depth $\lambda$ is assumed isotropic and homogeneous, $\theta$ is the quantum mechanical phase, $\vec{A}$ is the magnetic vector potential, and $\Phi_0=h/2e$ is the superconducting flux quantum. Taking the curl, the magnetic flux density $\vec{B}$ satisfies \cite{bluhm2007magnetic}
\begin{equation}
\nabla \times \nabla \times \vec{B}+\vec{B}/\lambda^2=\mu_0 \nabla\times \nabla \times \vec{M}.
\end{equation}
 We here neglect a term which represents a sum of delta function vortices. If we consider the case where there is a small susceptibility such that $\vec{M}=\chi\vec{H}$, and write $\bar{\mu}_2\equiv \mu_2/\mu_0=1+\chi_2$, with $\chi_2<<1$,  we then recover the familiar London's equation
\begin{equation}
\tilde{\lambda}^2\nabla^2 \vec{B}-\vec{B}=0,
\label{eq:london}
\end{equation}
with a modified penetration depth $\tilde{\lambda} \equiv \lambda/\sqrt{\bar{\mu}_2}. \,$\cite{matsumoto1982parameters,gray1983ginzburg,buzdin1986antiferromagnetic}  For all the experiments reported here $|\bar{\mu}_2-1|<<1$: the fit values for $\chi_2 t$  (see, e.g. Fig. \ref{fig:plot_correlation_map_3dmu}a) are less than 6$\times 10^{-4} \mu$m. Even if the layer responsible for the paramagnetism is only 10nm thick, this would correspond to $\chi_2=\bar{\mu}_2-1 < 0.06$. 

The fields in the 3 spatial regions of interest can then be expanded in Fourier series as
\begin{eqnarray}
\varphi_1(\vec{r},z)=&\frac{1}{(2\pi)^2} \int d^2k (\varphi_s(\vec{k})e^{kz}+\varphi_{r,1}(\vec{k})e^{-kz})e^{i \vec{k}\cdot\vec{r}} \nonumber \\
\vec{H}_2 (\vec{r},z)=&\frac{1}{(2\pi)^2} \int d^2k \left [ \vec{H}^+(\vec{k})e^{qz}+\vec{H}^-(\vec{k}) e^{-qz} \right ] e^{i \vec{k}\cdot\vec{r}}  \nonumber \\
\varphi_3(\vec{r},z)=&\frac{1}{(2\pi)^2} \int d^2k\, \varphi_{r,3}(\vec{k}) e^{kz} e^{i\vec{k}\cdot\vec{r}} ,
\label{eq:fourier_series}
\end{eqnarray}
where $\varphi_s$ is the source potential due to currents in the susceptometer field coil, and $\varphi_{r,1}$, $\varphi_{r,3}$, $\vec{H}^+$ and $\vec{H}^-$  are response potentials and fields, $k=|\vec{k}|$, and $q=(k^2+\tilde{\lambda}^{-2})^{1/2}$. Applying the boundary conditions of continuity of the normal component of $\vec{B}$ and the tangential component of $\vec{H}$ at the interfaces $z=0$ and $z=-t$, as well as the requirement that $\nabla\cdot\vec{B}=0$ in region 2, leads to the solution

\begin{widetext}
\begin{equation}
\varphi_{r,1}(k)=\frac{-(q+k\bar{\mu}_2)(-k\bar{\mu}_2+q\bar{\mu}_3)+e^{2 q t}(q-k\bar{\mu}_2)(k\bar{\mu}_2+q\bar{\mu}_3)}{-(q-k\bar{\mu}_2)(-k\bar{\mu}_2+q\bar{\mu}_3)+e^{2qt}(q+k\bar{\mu}_2)(k\bar{\mu}_2+q\bar{\mu}_3)}\varphi_s(k)
\label{eq:response_field}
\end{equation}
\end{widetext}

Simplified versions of this expression are given in Table \ref{tab:analytical_limits} in five limiting cases: a) is a bulk, non-superconducting paramagnet, b) is a thin, non-superconducting paramagnet, c) is a bulk superconductor without paramagnetism, with penetration depth short relative to the sensor field coil radius  and height, d) is a bulk superconductor without paramagnetism, with penetration depth long relative to the field coil radius and height, and e) is a thin superconductor without paramagnetism. It is of interest to note that in three of the cases: bulk paramagnetic (a), thin paramagnetic (b), and bulk weak diamagnetic (d), the material property of interest, either the permeability $\mu_2$ or the penetration depth $\lambda$, is separable from a geometrical factor, independent of the form of the source potential. This means that the temperature dependence of the material property can be determined, aside from a multiplicative constant, without curve fitting, in these cases. For example, in the bulk weak diamagnetic case (Table \ref{tab:analytical_limits}d), the SQUID susceptibility is proportional to $\lambda^{-2}$, with a constant of proportionality that depends only on geometry, independent of the form of the source term, which should be independent of temperature. This is also true in the thin diamagnetic case (Table \ref{tab:analytical_limits}e) for sufficiently large Pearl lengths $\Lambda$.  $\lambda$ and the geometrical factors are not separable in the limiting case of bulk strong diamagnetism (Table \ref{tab:analytical_limits}c). In this case, to a good approximation (see Appendix \ref{sec:analytical_approx}) it is the sum $\lambda+z_0$ which is determined by SQUID susceptibility measurements.\cite{kogan2003meissner}

The source field for a circular field coil of radius $a$ is given by \cite{kogan2003meissner}
\begin{equation}
\varphi_s(k)=\frac{\pi I a}{k} e^{-k z} J_1(k a)
\label{eq:source}
\end{equation}
The $z$-component of the response field in region 1 is given by $h_r(k,z)=-k\varphi_{r,1}e^{-kz}$. Taking the limit $b<<a$,  the height dependence of the SQUID susceptibility $\phi(z)$ is given by

\begin{widetext}
\begin{equation}
\phi(z)/\phi_s = \int_0^{\infty} dx \, xe^{-2x\bar{z}} J_1(x) \left [ \frac{-(\bar{q}+\bar{\mu}_2 x)(\bar{\mu}_3\bar{q}-\bar{\mu}_2x)+e^{2\bar{q}\bar{t}}(\bar{q}-\bar{\mu}_2x)(\bar{\mu}_3\bar{q}+\bar{\mu}_2x)}{-(\bar{q}-\bar{\mu}_2x)(\bar{\mu}_3\bar{q}-\bar{\mu}_2x)+e^{2\bar{q}\bar{t}}(\bar{q}+\bar{\mu}_2x)(\bar{\mu}_3\bar{q}+\bar{\mu}_2x)} \right ],
\label{eq:full_susc_expression}
\end{equation}
\end{widetext}

where 
\begin{equation}
\phi \equiv \frac{1}{\Phi_0}\frac{d\Phi}{dI}, 
\label{eq:phi0}
\end{equation}

$\Phi$ is the flux through the pickup loop in response to the current $I$,  $\Phi_0=h/2e$ is the superconducting flux quantum, the self inductance between the field coil and the pickup loop 
\begin{equation}
\phi_s=A\mu_0/2\Phi_0 a,
\label{eq:phi_0}
\end{equation}  
$A$ is the effective area of the pickup loop, $\bar{z}=z/a$, $\bar{t}=t/a$, and $\bar{q}=\sqrt{x^2+1/\bar{\lambda}^2}$, with $\bar{\lambda} \equiv \tilde{\lambda}/a$.

\subsection{Analytical expressions in various limits}

\begin{table*}[ht]
\begin{tabular}{| c | c | c | c | c | c |c|}
\hline
& Description & Thickness & Penetration depth & Permeability & $\varphi_{r,1}(k)/\varphi_s(k)$& $\phi(z)/\phi_s$ \\
\hline
a & Bulk para                    &   $\bar{t}>>1,\bar{z}$      &   $\bar{\lambda} >> 1,\bar{z}$       & $\bar{\mu}_2 > 1$                           &$-\frac{\bar{\mu}_2+1}{\bar{\mu}_2+1} $        &     $\left ( \frac{\bar{\mu}_2-1}{\bar{\mu}_2+1} \right ) \frac{1}{(1+4\bar{z}^2)^{3/2}}$ \\
b & Thin para                    & $\bar{t}<<1,\bar{z}$        &   $\bar{\lambda} >>1,\bar{z}$        &$\bar{\mu}_2>1, \bar{\mu}_3=1$  &$-\frac{kt(\bar{\mu}_2^2-1)}{2\bar{\mu}_2}$  & $   \frac{\bar{\mu}_2^2-1}{\bar{\mu}_2} \frac{3\bar{t}\bar{z}}{(1+4\bar{z}^2)^{5/2}}$ \\
c & Bulk strong dia           &    $\bar{t}>>1$                 &    $\bar{\lambda} << 1,\bar{z}$      & $\bar{\mu}_2 = 1$                           & $\frac{1}{\lambda^2(q+k)^2}   $                     &     $-\frac{1}{(1+4(\bar{z}+\bar{\lambda})^2)^{3/2}}$ \\
d & Bulk weak dia            & $\bar{t}>>1,\bar{z}$        &   $\bar{\lambda} >> 1,\bar{z}  $     & $\bar{\mu}_2=\bar{\mu}_3=1$     &$\frac{1}{4\lambda^2k^2}$                               & $-\frac{1}{4\bar{\lambda}^2}(\sqrt{4\bar{z}^2+1}-2\bar{z})$ \\
e & Thin dia                       &   $\bar{t}<<1,\bar{z}$      &   $\bar{\lambda} >> \bar{t} $          &$\bar{\mu}_2=\bar{\mu}_3=1$      &$\frac{1}{1+\Lambda k}$                                   & $-\frac{a}{\Lambda} \left ( 1-\frac{2\bar{z}}{\sqrt{1+4\bar{z}^2}} \right ) $ \\
\hline

\end{tabular}
\caption{Response scalar potential $\varphi_{r,1}(k)$ divided by the source potential $\varphi_s(k)$ in momentum space (Eq. \ref{eq:response_field}) and scanning SQUID susceptibility $\phi(z)$ divided by the SQUID self-susceptibility $\phi_s$ in real space (Eq. \ref{eq:full_susc_expression}), in various limits.}
\label{tab:analytical_limits}
\end{table*}

In general the integral in Eq. \ref{eq:full_susc_expression} must be done numerically, but analytical expressions can be derived in the limits given in Table \ref{tab:analytical_limits} and plotted in Figure \ref{fig:plot_susc_limits}:  
\begin{figure}
\includegraphics[width=3.3in,trim=0 0 0 0]{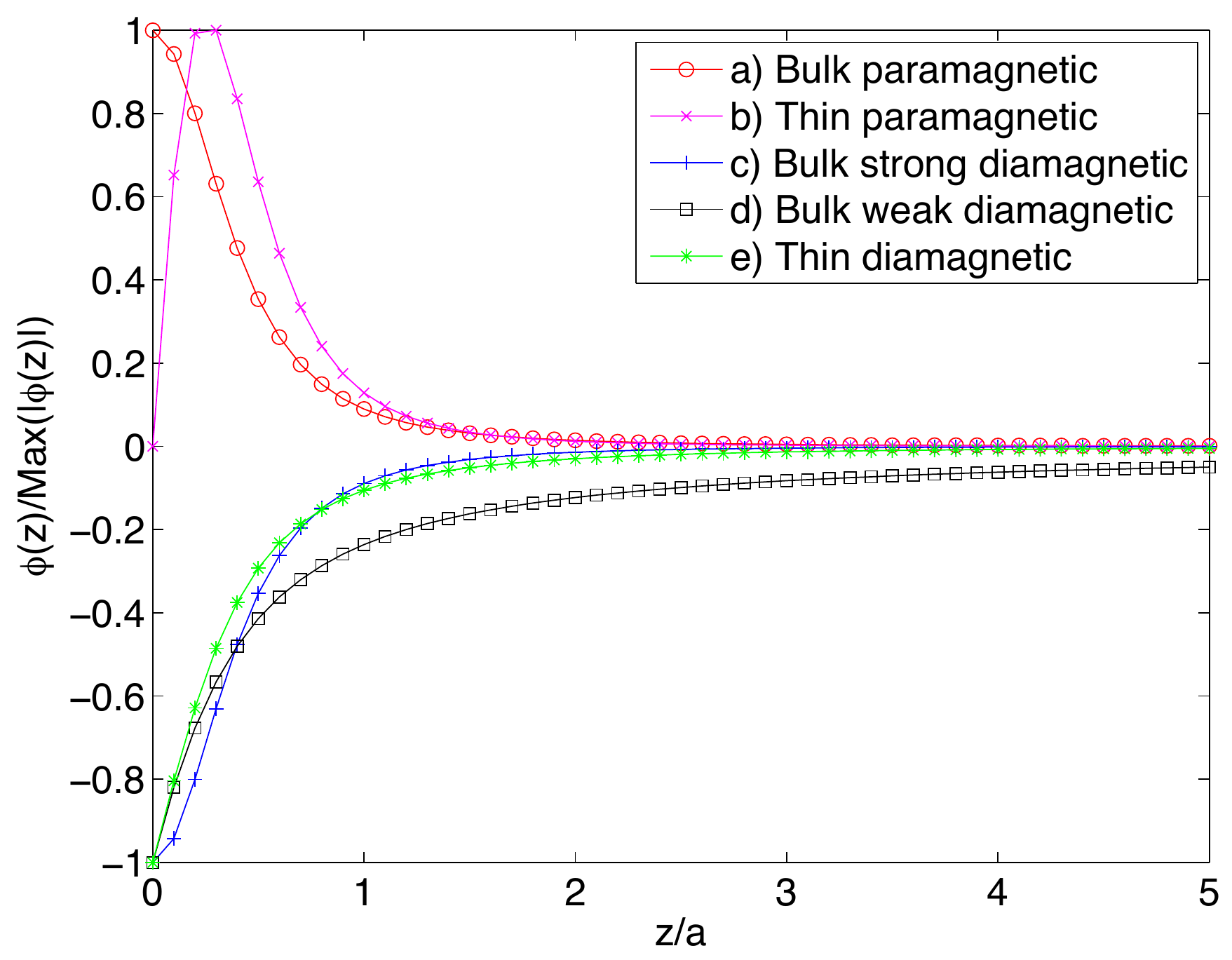}
\caption{Theoretical height dependence of the scanning SQUID susceptibility, divided by the maximum of the absolute value of the susceptibility, of a paramagnetic superconductor in various limits. The letters correspond to the entries in Table \ref{tab:analytical_limits}.}
\label{fig:plot_susc_limits}
\end{figure}

It is to be noted that in the bulk limit ($\bar{t} >>1$) the strong superconducting susceptibility ($\bar{\lambda} <<1$, $\bar{\mu}_2=1$) and the paramagnetic susceptibility ($\bar{\lambda}>>1$, $\bar{\mu}_2 > 1$) have the same height dependence aside from a scaling factor $-(\bar{\mu}_2-1)/(\bar{\mu}_2+1)$. However, when the thickness of the paramagnetic superconductor becomes comparable to the field coil radius, the height dependences of the paramagnetic and superconducting contributions become different, and it is possible in principle to determine the relative contributions of each to the total susceptibility by fitting approach curves. It is also possible in principle to determine the $z$-dependence of the response carrier density (either paramagnetic or diamagnetic) from approach curves. However, in practice the differences between the spacing dependences of the various contributions are subtle, and it is difficult to separate out the paramagnetic from the diamagnetic components without extra information. Such information could be supplied, for example, by raising the temperature above the superconducting transition temperature, leaving only the paramagnetic contribution, or studying the low temperature dependence of the susceptibility, where the temperature dependence of the superconducting component could saturate, while that of the paramagnetic component could become larger. Finally, one or both components could be spatially dependent (see e.g. Fig. \ref{fig:2DEL_images}), which could help to separate them. The regions in parameter space of validity  and errors associated with using the approximate expressions in Table \ref{tab:analytical_limits} are explored in Appendix \ref{sec:analytical_approx}.



\section{Comparison with experiments}

\begin{figure}
\includegraphics[width=3.3in,trim=0 0 0 0]{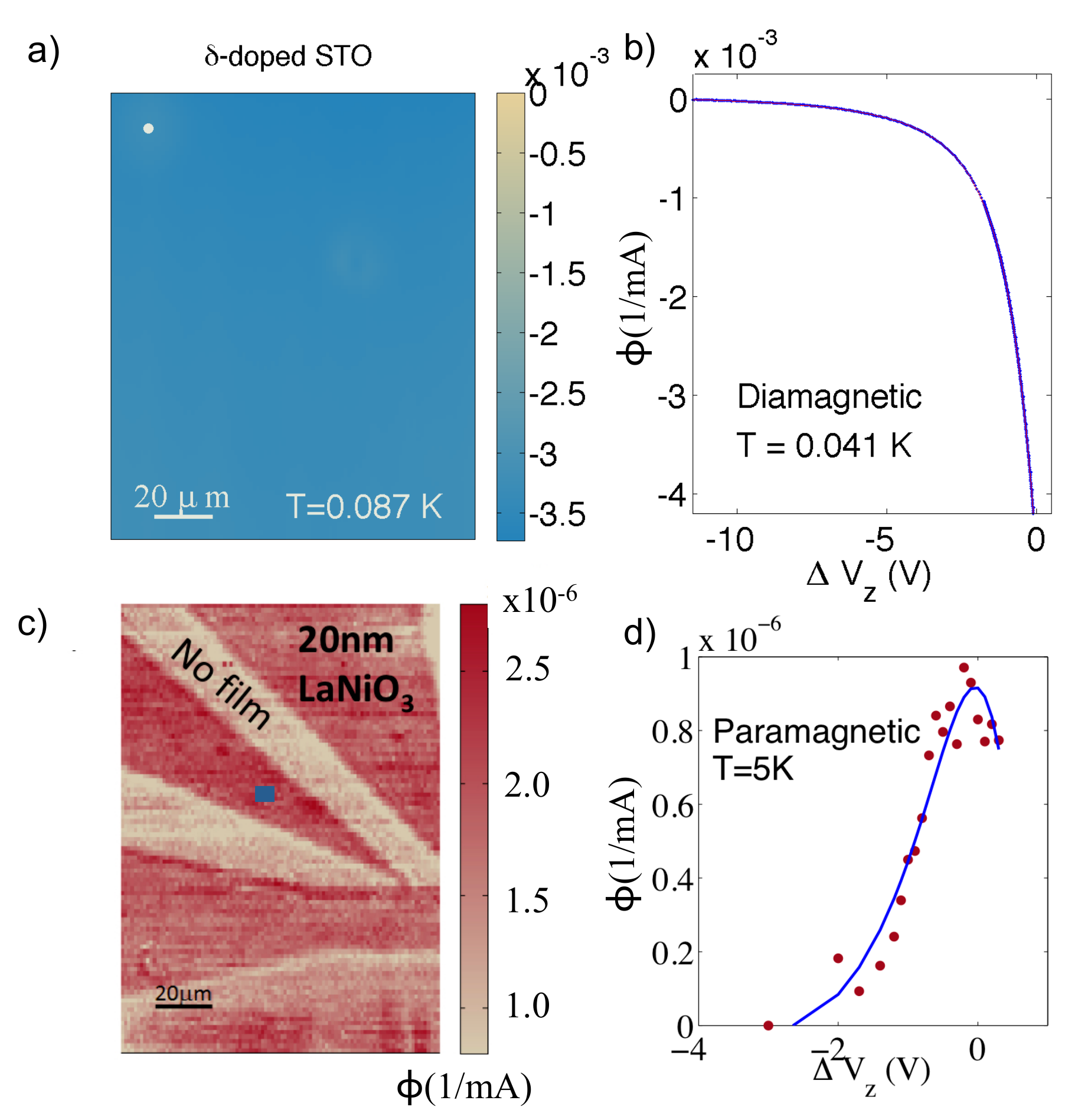}
\caption{ a) Susceptometry image of a $\delta$-doped SrTiO$_3$ sample.
b) Susceptibility as a function of $\Delta V_z$, the change in z-piezo voltage from contact between the SQUID substrate and the sample surface, at the position of the gray circle in a). The dots are data, the line is a fit using the thin diamagnetic limit expression in Table \ref{tab:analytical_limits}(e), with $\Lambda=954\mu m$.
c) Scanning susceptibility image of a patterned non-superconducting thin film of LaNiO$_3$.
d) Susceptibility approach curve for the LaNiO$_3$ film at the position of the square symbol in (c). The dots are data, the line is a fit using the thin paramagnetic limit expression (Table \ref{tab:analytical_limits}(b)), with $\chi_2 t=1.3\times10^{-5}\mu m$.
}
\label{fig:introduction_figure4}
\end{figure}

There have been a number of works in which SQUID susceptibility measurements have been used to infer the London penetration depth of superconductors.\cite{gardner2001scanning,tafuri2004magnetic,hicks2009evidence,luan2011local,kalisky2010stripes} We examine scanning SQUID data from several samples. The low temperature measurements were performed in a home-built SQUID microscope in a dilution refrigerator.\cite{bjornsson2001scanning} The 5K measurements were performed in a home built variable sample temperature scanning SQUID microscope.\cite{kalisky2010stripes} The SQUID susceptometers used in both microscopes were described in Ref. \onlinecite{huber2008gradiometric}.

Figure \ref{fig:introduction_figure4} shows experimental data for samples with predominantly diamagnetic response (Fig. \ref{fig:introduction_figure4}a,b) and paramagnetic response (Fig. \ref{fig:introduction_figure4}c,d). Fig. \ref{fig:introduction_figure4} (a) and (b) show SQUID susceptometry of a Nb $\delta$-doped sample of SrTiO$_3$(STO).\cite{bert2011direct} This sample was grown in an atmosphere of $10^{-8}$ Torr oxygen at 1200$^\circ$C. Nb dopants were confined to a 5.9nm layer, with 100 nm cap and buffer layers of STO grown above and below the doped region. The sample was annealed {\it in situ} under an oxygen partial pressure of $10^{-2}$Torr at 900$^{\circ}$C for 30 minutes.\cite{kozuka2009two}
For the data sets of Fig. \ref{fig:introduction_figure4} (b) and (d), the susceptibility was recorded while the SQUID was driven towards the sample by increasing the z-piezo voltage $V_z$ from a large negative value. In these plots $\Delta V_z=0$ corresponds to contact between the SQUID substrate and the sample surface. At contact the SQUID pickup loop is a height $z_0$ above the sample surface because of the finite angle (typically a few degrees) between the sample surface and the SQUID substrate surface. The dots in Fig. \ref{fig:introduction_figure4}(b) show such a susceptometry approach curve at the position indicated by the gray circle in Fig. \ref{fig:introduction_figure4}(a). The line (difficult to distinguish from the data in this plot) is a fit of this data to the thin diamagnetic limit of Table \ref{tab:analytical_limits} with 5 fitting parameters: a vertical shift $\delta\phi$, a linear slope $\phi_{\rm linear} = \alpha z$, the Pearl length $\Lambda$, $z_0$, and the change in $z$ with piezo voltage $dz/dV_z$. We do not know the source of the linear background. In the present case it was small, $\alpha \sim -3.7\times10^{-6}$ 1/mA-$\mu$m. 

The two fixed parameters in this analysis were the effective field coil radius $a$=8.4$\mu$m, and pickup loop radius $b$=2.7$\mu$m. The effective field coil radius was taken from the numerical calculations of Brandt and Clem,\cite{brandt2004superconducting} using (see Fig. \ref{fig:slab_susc_geo}a) a field coil inside radius of 6.5$\mu$m, outside radius of 12$\mu$m, thickness 0.3$\mu$m, and penetration depth 0.09$\mu$m. The effective pickup loop radius was chosen to result in the measured self inductance of $\phi_s$ = 800 1/A using Eq. \ref{eq:phi_0}. This results in an effective pickup loop area of 22$\mu$m$^2$, larger than the 17$\mu$m$^2$ obtained from the sum of the geometric mean of the pickup loop itself, with inside radius $r_{in}$=0.88$\mu$m, and outside radius $r_{out}$=2.4$\mu$m, added to the Ketchen's 1/3 rule area (w$^2$/3)[\onlinecite{ketchen1995design}] for the shield over the pickup loop leads, which has $w$=4.5$\mu$m. Part of this discrepancy may be due to the fact that the pickup loop shield focusses flux from the field coil into the pickup loop area. 

\begin{figure}
\includegraphics[width=3in,trim=0 0 0 0]{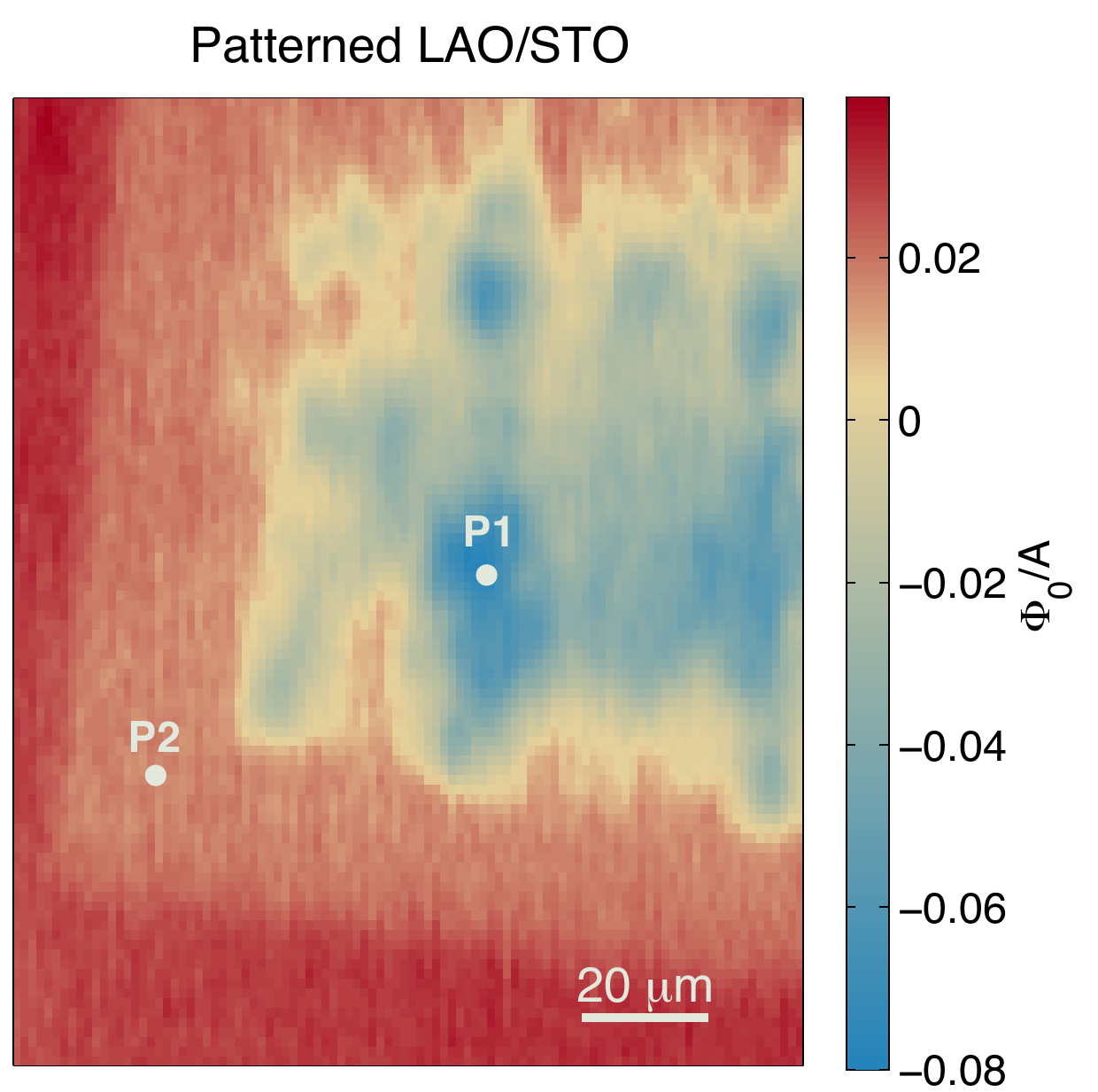}
\caption{Scanning SQUID susceptometry image of a patterned LAO/STO interface at 0.087K. The labels indicate where the data in Fig. \ref{fig:plot_susc_fits} was taken.}
\label{fig:2DEL_images}
\end{figure}

\begin{figure}
\includegraphics[width=3in,trim=0 0 0 0]{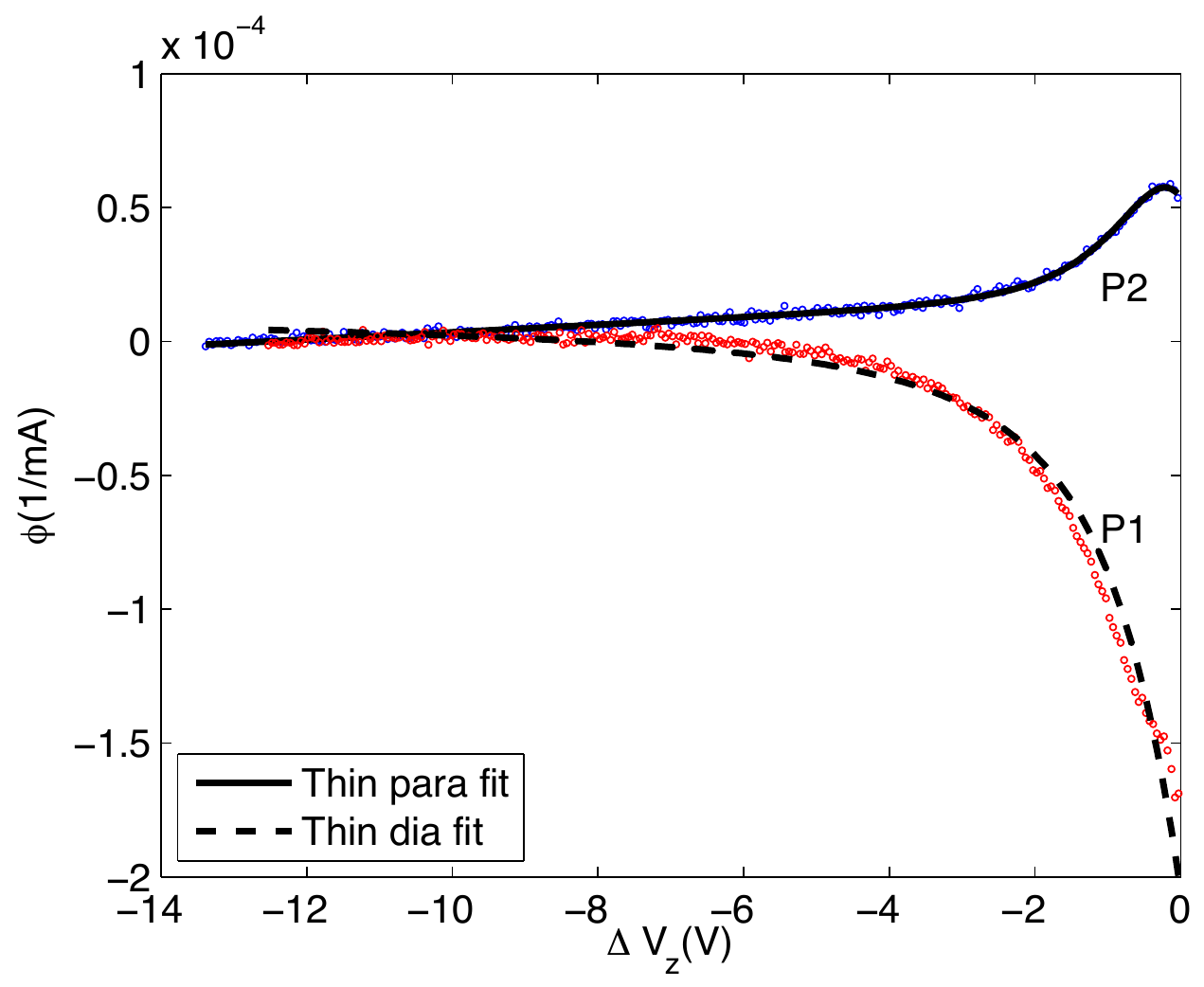}
\caption{Scanning SQUID susceptibility as a function of sensor-sample spacing at 2 positions on the patterned LAO/STO sample of Fig. \ref{fig:2DEL_images}: P1, in a region showing predominantly diamagnetic shielding; and P2, in a region showing predominantly paramagnetic response. The symbols are data and the lines are fits as described in the text.
}
\label{fig:plot_susc_fits}
\end{figure}


The paramagnetic susceptibility of samples is typically much smaller than the diamagnetic susceptibility of superconductors. An example is shown in Fig. \ref{fig:introduction_figure4}(c) and (d). Fig. \ref{fig:introduction_figure4}(c) shows a scanning susceptometry image of a patterned 20nm thick film of LaNiO$_3$, imaged at 5K with a SQUID susceptometer with the same geometry as in Fig. \ref{fig:introduction_figure4}(a). For this sample thermally evaporated La and Ni were co-deposited on to a LaAlO$_3$ substrate kept at a temperature of 600C, in a background oxygen pressure of $7\times10^{-6}$ Torr.  An RF source operated at 200 W provided atomic oxygen to the film during growth, and in situ structural characterization was obtained using RHEED. Patterning of the film was achieved by creating a mask on the film surface using photolithography, and then etching the film in a HCl solution (4:1 H2O:HCl) to remove uncovered areas. The dots in Fig. \ref{fig:introduction_figure4}d are the data, the line is a fit to the thin paramagnetic limit expression of Table \ref{tab:analytical_limits}(b), with 4 fitting parameters: $\delta\phi$, $\chi_2 t=1.3\times10^{-5}\mu$m, $z_0$=1.9$\mu$m, and $dz/dV_z$=2.7$\mu$m/V, where $\chi_2=\mu_2/\mu_0-1$ and $t$ is the thickness of the film. 

As an example of a sample that shows both paramagnetic and diamagnetic behavior we present data on the 2-DEL at the interface between the perovskite insulators SrTiO$_3$ (STO) and LaAlO$_3$ (LAO). The sample and measurement techniques for the data used in this study were described in Ref. \onlinecite{bert2011direct}. Briefly, the sample was prepared by growing 10 unit cells of LaAlO$_3$ on a commercial TiO$_2$ -terminated 001 STO substrate, with an aluminum oxide hard mask patterned on to the STO substrate prior to LAO growth.  A crystalline LAO/STO interface only grew in the gaps of the patterned mask. The LaAlO$_3$ was deposited at 800$^\circ$ C with an oxygen partial pressure of 10$^{-5}$ mbar, after a pre-anneal at 950$^\circ$ C with an oxygen partial pressure of $5\times 10^{-6}$ mbar for 30 min. The sample was cooled to 600$^\circ$ C and annealed in a high-pressure oxygen environment (0.4 bar) for one hour. Figure \ref{fig:plot_susc_fits} displays SQUID susceptibility data as a function of spacing between the sensor and the sample surface for the LAO/STO sample imaged in Fig. \ref{fig:2DEL_images} at the positions labelled. Both positions are in a gap of the aluminum oxide mask, but P2, close to the edge of the two-dimensional electron layer, shows paramagnetic response, while P1 shows diamagnetic behavior. These approach curve data were taken at T=0.02K with a field coil current of 1mA.

 
 The fact that P2 shows a maximum below $\Delta V_z=0$ implies that the paramagnetism results from a thin film, rather than from the substrate (compare the thin and bulk paramagnetic limit curves in Fig. \ref{fig:plot_susc_limits}). Fitting this data to the pure thin paramagnetic expression of Table \ref{tab:analytical_limits}, with $\chi_2t$, $z_0$, $\alpha$ and $\delta\phi$ as variables, with $a$=8.4$\mu$m, $b$=2.7$\mu$m and $dz/dV=2.9\mu m/V$, results in $\chi_2t$=4.9+0.8-0.7$\times 10^{-4}\mu m$ and $z_0=1.5+0.7-0.3\mu m$. This fit is displayed as the solid line in Fig. \ref{fig:plot_susc_fits}. Using the same assumptions as for the LaNiO$_3$ case above (including a $\pm$20\% uncertainty in $a$), we find a spin density of 1.25$\pm$0.5$\times 10^{14}$1/cm$^2$. 

We attempted to fit curve P1 in Figure \ref{fig:plot_susc_fits} to the pure thin film diamagnetic expression of Table \ref{tab:analytical_limits}(e), with $\Lambda$, $z_0$, $\alpha$, and $\delta\phi$ as variables, and $dz/dV=2.9\mu m/V$ as a fixed parameter. The best fits were obtained for unphysical negative values for $z_0$. If we constrain $z_0$ to vary between the values of 1 and 2.5$\mu$m,  the best fit (dashed line in Fig. \ref{fig:plot_susc_fits}) occurs for $z_0=2.5\mu$m and $\Lambda=16.4\mu m$. However, the fit quality was not good (the best fit $\Xi^2_{\rm min}$ is about 25 times worse for P1 than for P2). Using the same assumptions as for $\delta$-STO above, but with an effective mass $m*=1.46m_e$\cite{caviglia2010two}, the allowed values for $\Lambda$ (15 mm $< \Lambda <$ 34mm) correspond to a Cooper pair density of 1$\times 10^{11}$cm$^{-2} < N_s < 3.4\times 10^{11}$cm$^{-2}$.

It seems reasonable to assume that the susceptibility at position P1 in Fig. \ref{fig:2DEL_images} has both superconducting and paramagnetic contributions, and therefore could be fit using the full expression Eq. \ref{eq:full_susc_expression}. However, the fitting parameters $\chi_2t$, $z_0$ and $\Lambda$ are strongly correlated, resulting in large uncertainties in their values. Further, fits to this data result in unphysical negative best fit values for $\chi_2t$ and $z_0$. We speculate that these unphysical values might result from the inhomogeneous superfluid density in this sample or from interaction between the sensor SQUID and the superfluid at these low densities. Therefore, as mentioned in the introduction, additional information, such as different temperature or spatial dependences, will be required to separate the superconducting from the paramagnetic components in scanning SQUID susceptibility measurements.

\section{Conclusions}

We have presented a full expression and analytical approximations in various limits for the susceptibility in a scanning SQUID geometry of a paramagnetic superconductor of arbitrary thickness on a paramagnetic substrate. These expressions can be used to measure the spin concentration and the Cooper pair density in a paramagnetic superconductor. A comparison of $\Xi^2$ analysis with bootstrap statistical analyses (see e.g. Appendix \ref{sec:param_uncertainties}, Fig. \ref{fig:plot_correlation_map_3d}) indicate that the accuracy of these measurements can be improved  with a precise knowledge of the sensor height $z_0$ and the piezo constant $dz/dV$ in scanning SQUID susceptometry measurements.

\section{Acknowledgements}

The experimental work on LAO/STO was  supported by the US Department of Energy, Office of Basic Energy Sciences, Division of Materials Sciences and Engineering under award DE-AC02-76SF00515. J.K. and B.K. were supported by the NSF under Grant No. PHY-0425897. J.H.N, Y.S., F.J.W. and C.H.A. were supported by a DARPA COMPASS grant. We would like to thank Vladimir Kogan for his comments and suggestions.

\appendix*

\section{Sources of error}

\subsection{Systematic}

\subsubsection{Approximating field coil by circular wire}
\label{sec:field_coil_approx}
The actual susceptometer layout used in the experiments described in this paper is shown in Figure \ref{fig:slab_susc_geo}a. A full calculation of the fields generated by the field coil would require a three-dimensional solution of coupled London's and Maxwell's equations in this geometry. To estimate the errors associated with approximating the actual field coil geometry by an infinitely thin circular wire, we consider the idealized geometry of Fig. \ref{fig:slab_susc_geo}b: the field coil is assumed to be an incomplete, infinitely narrow circle of radius $a$, which connects with infinitely long, infinitely narrow leads with spacing $s$, and which carries a current $I$. For these calculations, we take $a$=8.4$\mu$m and $s=7.3\mu m$. $s$ was taken as the geometric mean of the outside $13\mu m$ and inside $1.2\mu m$ widths of the leads in the susceptometer layout.

\begin{figure}
\includegraphics[width=3.3in,trim=0 0 0 0]{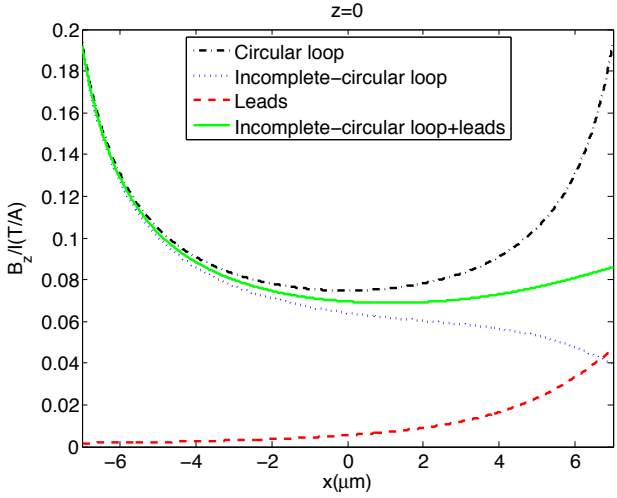}
\caption{Calculated $z$-component of the field $B_z$, divided by the current $I$ for the model of Fig. \ref{fig:slab_susc_geo}b, for a circular loop of the same radius (dot-dashed curve), the contribution from the incomplete circular loop (dotted curve), the contribution from the leads (dashed curve) and the sum of the previous two (solid line). The $x$-axis is oriented along the leads, with the leads coming in along the positive $x$-direction. This calculation assumes the pickup loop and the field coil are in the same plane. The field at the center is  7.5\% larger for the circular loop model than for the incomplete circle plus leads model.}
\label{fig:systematic_curves}
\end{figure}

Using Biot-Savart:
\begin{equation}
\vec{B}=\frac{\mu_0 I}{4\pi} \oint \frac{\vec{dl}\times\vec{r}}{| \vec{r} |^3}
\label{eq:biot}
\end{equation}
the contribution $B_{z,c}$  to the $z$-component of the field from the incomplete loop is given by
\begin{widetext}
\begin{equation}
B_{z,c} = \frac{\mu_o I}{4\pi} \int_{\theta_1}^{2\pi-\theta_1} \, d\theta \frac{a^2-ay\sin \theta - a x \cos \theta}{((x-a\cos \theta)^2+(y-a \sin \theta)^2+z^2)^{3/2}},
\label{eq:circ}
\end{equation}
where $\theta_1=\cos^{-1}(s/2a)$.
The contribution $B_{z,l}$ from the leads is
\begin{equation}
B_{z,l}=\frac{\mu_0 I}{2\pi} \frac{2(s-2y)(2(x-x_0)+\sqrt{(s-2y)^2+4((x-x_0)^2+z^2)})}{((s-2y)^2+4z^2)\sqrt{(s-2y)^2+4((x-x_0)^2+z^2)}},
\label{eq:leads}
\end{equation}
\end{widetext}
where $x_0=a\cos \theta_1$. Figure \ref{fig:systematic_curves} plots the $z$-components of the fields from a circular loop ($\theta_1=0$), from an incomplete circle, from the leads, and the sum of the incomplete circle and leads, using the parameters above, and assuming $z$=0: the field coil and the pickup loop in the same plane. The field from the circular coil model is 7.5\% higher than that from the incomplete circle plus leads model at $z=0$. At a height of $z=3\mu m$, more appropriate for calculating a susceptibility, the error is 6.8\%.

\begin{figure}
\includegraphics[width=3.3in,trim=0 0 0 0]{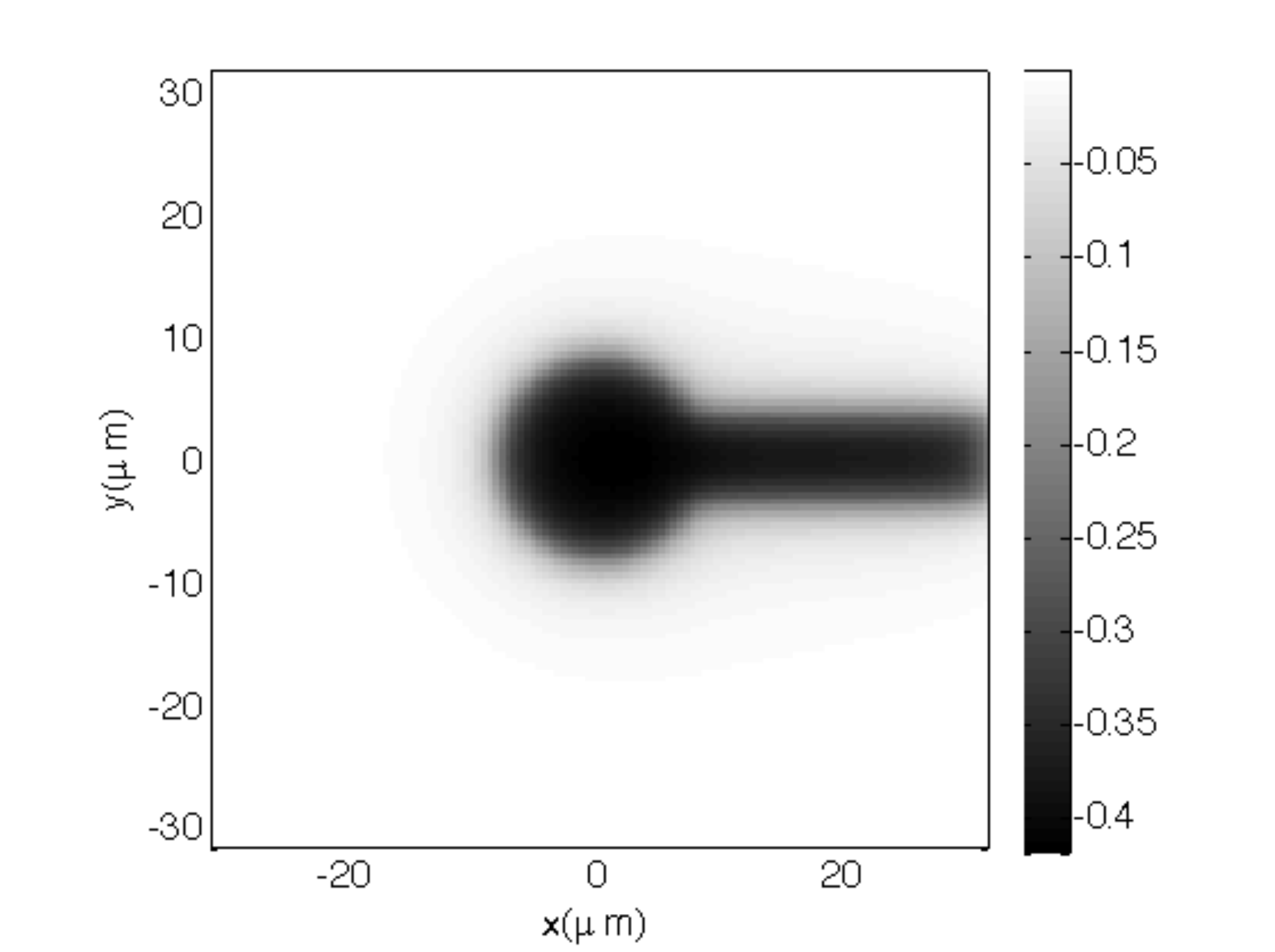}

\caption{Calculated magnetic scalar potential $\varphi_s(\vec{r},0)$, divided by the current through the loop, for the model of Fig. \ref{fig:slab_susc_geo}c with $a$=8.4$\mu m$, $s=7.4\mu m$ and $z_0=1.5\mu m$.}
\label{fig:systematic_scalar_potential}
\end{figure}

To calculate the SQUID susceptibility using the present formalism requires calculating the magnetic scalar source potential $\varphi_s(\vec{r},0)$ for the above geometry. Converting Eq. 36 of Reference \onlinecite{kogan2003meissner} to SI units:
\begin{equation}
\varphi_s(\vec{r},0)=\frac{Iz_0}{4\pi} \int \frac{d^2 \vec{r'}}{((\vec{r}-\vec{r'})^2+z_0^2)^{3/2}}
\label{eq:magnetic_scalar_potential}
\end{equation}
This is difficult to integrate over an arbitrary geometry analytically. Instead we did the integrations and Fourier transforms numerically. Figure \ref{fig:systematic_scalar_potential} shows the results for $\varphi_s(\vec{r},0)/I$ (which is dimensionless) for $a=8.4\mu m$, $s=7.3\mu m$ and $z_0=1.5\mu m$. 

\begin{figure}
\includegraphics[width=3.3in,trim=0 0 0 0]{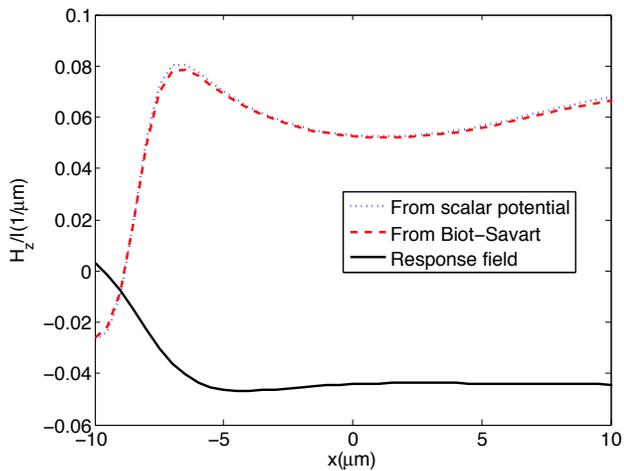}
\caption{ $z$-component of the source field, divided by the current,  with $a$=8.4$\mu m$, $s=7.4\mu m$ and $z_0=1.5\mu m$, calculated using Biot-Savart vs taking the gradient of the scalar potential. Also shown for comparison is a cross-section through the response field image of Fig. \ref{fig:systematic_response_field}.}
\label{fig:systematic_field_crosses}
\end{figure}

The source field is given by $\vec{H_s}=\vec{\nabla}\varphi_s(\vec{r},z)$. Figure \ref{fig:systematic_field_crosses} compares the results obtained using Biot-Savart with the gradient of the scalar potential of Fig. \ref{fig:systematic_scalar_potential}.  

\begin{figure}
\includegraphics[width=3.3in,trim=0 0 0 0]{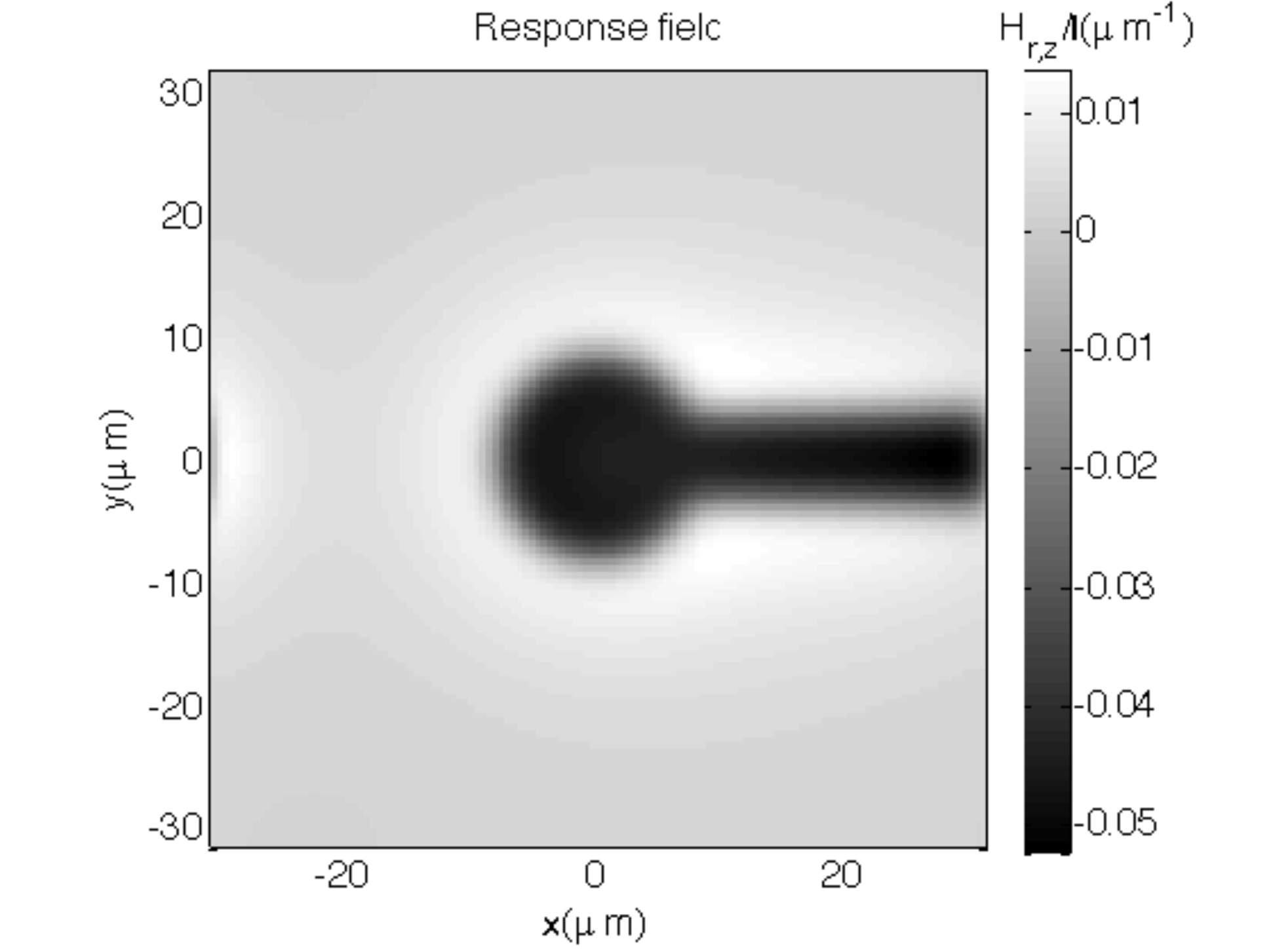}
\caption{ $z$-component of the response field, divided by the current, with $a$=8.4$\mu m$, $s=7.4\mu m$ and $z_0=1.5\mu m$, t=10$\mu m$, and $\lambda=0.1\mu m$.}
\label{fig:systematic_response_field}
\end{figure}

The Fourier transform of the response field at $z=z_0$ is given by
\begin{equation}
h_{z,r}(\vec{k},z_0)=-k \varphi_{r,1}({k}) e^{-kz_0}
\label{eq:h_z}
\end{equation}
where $\varphi_{r,1}(k)$ is given by Eq. \ref{eq:response_field}.
Figure \ref{fig:systematic_response_field} shows the results for the response field using $a$=8.4$\mu m$, $s=7.4\mu m$ and $z_0=1.5\mu m$, a sample thickness of $t=10\mu m$, $\lambda=0.1\mu m$, $\mu_2/\mu_0=1$, $\mu_3/\mu_0=1$: the strong diamagnetic shielding, bulk, non-paramagnetic limit of Eq. \ref{eq:response_field}. Table \ref{tab:systematic_calculations} shows some selected comparisons of calculations using the circular wire model with the incomplete circle plus leads model for the field coil.


\begin{table*}[ht]
\begin{tabular}{| c | c | c | c | c | c | c | c | c | c |}
\hline
$a(\mu m)$ & $z_0(\mu m)$ & $s(\mu m)$ & $t(\mu m)$ & $\lambda(\mu m)$ & $ \frac{H_{s,z}(0)}{I} (\mu m^{-1}) $ & $\frac{1}{2a} (\mu m^{-1})$ & $\frac{H_{r,z}(z_0)}{I} (\mu m^{-1}) $ & $\frac{H_{r,z}(z_0)}{H_{s,z}(0)}$ & Analytic limit \\
\hline
8.4 & 1.5 & 7.3 & 0.1 & 10 & $5.27\times10^{-2}$ & $ 5.95\times10^{-2}$ & $1.44\times10^{-4}$ & $2.74\times10^{-3}$ & $2.70\times10^{-3}$ \\
12 & 1.5 & 13 & 0.1 & 10 & $3.71\times10^{-2}$ & $ 4.17\times10^{-2}$ & $1.50\times10^{-4}$ & $4.05\times10^{-3}$ & $4.45\times10^{-3}$ \\
6   & 1.5 & 1.2  & 0.1 & 10 & $7.41\times10^{-2}$ & $ 8.33\times10^{-2}$ & $1.26\times10^{-4}$ & $1.69\times10^{-3}$ & $1.59\times10^{-3}$ \\
8.4 & 1.5 & 7.3 & 10 & 0.1 & $5.27\times10^{-2}$ & $ 5.95\times10^{-2}$ & $4.42\times10^{-3}$ & $0.836$ & $0.816$ \\
12  & 1.5 & 13 & 10 & 0.1 & $3.71\times10^{-2}$ & $ 4.17\times10^{-2}$ & $3.27\times10^{-2}$ & $0.882$ & $0.902$ \\
6    & 1.5 & 1.2 & 10 & 0.1 & $7.41\times10^{-2}$ & $ 8.33\times10^{-2}$ & $5.50\times10^{-2}$ & $0.742$ & $0.687$ \\

\hline

\end{tabular}
\caption{Some results from the evaluation of Eq.'s \ref{eq:magnetic_scalar_potential}, \ref{eq:h_z}, and \ref{eq:response_field} for various parameters. The first 3 rows are in the thin film diamagnetic limit, and the last 3 are in the strong bulk diamagnetic limit. }
\label{tab:systematic_calculations}
\end{table*}

The first 5 columns of Table \ref{tab:systematic_calculations} are  the field coil radius $a$, the height of the susceptometer above the sample surface $z_0$,  the spacing between the leads $s$, the thickness of the paramagnetic superconductor $t$ ,  and the London penetration depth  $\lambda$. $H_z(0)/I$ is the field at the center of the coil divided by the current through the coil. This is approximated in the circular wire model for the field coil by $1/2a$. $H_{zr}(z_0)/I$ is the response field at the center of the field coil (and pickup loop), divided by the current through the field coil. Finally $H_{zr}(z_0)/H_z(0)$ is equivalent to $\phi(z)/\phi_s$, the ratio of the sample susceptibility to the self-inductance in the limit where the pickup loop radius $b<<a$. In all cases $\mu_2=\mu_3=\mu_0$ in Table \ref{tab:systematic_calculations}.

In the thin diamagnetic limit (first 3 rows of the table) $\phi(z)/\phi_s \rightarrow -(a/\Lambda)(1-2(z/a)/\sqrt{1+4(z/a)^2})$. In the bulk diamagnetic limit $\phi(z)/\phi_s \rightarrow -1/(1+4(z+\lambda)^2/a^2)^{3/2}$. These values are entered in the last column of the table. A comparison of the last 2 columns of the table shows that if one normalizes by the self-susceptibility, the analytic limits derived above for the circular field coil model agree with the full incomplete circle with leads model to within 10\%,  independent of whether the current is localized at the very inside of the field coil or at the outside of the field coil, and presumably for any current distribution in between.

A more rigorous solution of the problem would solve London's equations for the current distribution in the field coil following, e.g. Brandt and Clem, \cite{brandt2004superconducting}  then use those results to find the scalar potentials for a set of equi-current-density paths, and add them up with suitable weightings. However, the results of Table \ref{tab:systematic_calculations} indicate that the results of such a complex calculation would not differ from the infinitely narrow, circular field coil model by more than 10 percent.

\subsubsection{Approximating flux in pickup loop by field at center of field coil times an effective area}
\label{sec:pickup_loop_approx}

A simplification used in this paper is to approximate the flux through the pickup loop by the field at the center of the field coil times an effective area. More traditional\cite{ketchen1995design} is to model the pickup loop area as composed of a circle of radius $b$, co-planar and concentric with the field coil. An additional pickup area from the leads is approximated by a square of width and length $w$, offset from the center of the pickup loop circle by a length $\Delta w$. The square area contributes one third of the flux passing through it to the total pickup loop flux. For these calculations, we take $a$=8.4$\mu$m, $s=5\mu m$, $b$=1.8$\mu$m, $w$=4.5$\mu$m and $\Delta w$=1.5$\mu$m.

Numerical integration of the field from the field coil  gives an integrated flux through the pickup loop (690$\Phi_0$/A) that is 11\% larger than the flux $B_z(0)A$ (612$\Phi_0$/A), where $B_z(0)$ is the field at the center of the field coil, and $A=\pi b^2+w^2/3$ is the effective area of the pickup loop in the circular field coil model (see Appendix \ref{sec:field_coil_approx}), and 16\% larger (693$\Phi_0$/A vs. 583$\Phi_0$/A) in the incomplete circle plus leads model, assuming $z=0$. This would be appropriate for calculating the self-inductance of the susceptometer, and could help to explain why it is necessary to use a somewhat larger effective area for the pickup loop (22$\mu m^2$ rather than 17$\mu m^2$) than the Ketchen model\cite{ketchen1995design} gives. More appropriate for estimating uncertainties in susceptibility measurements would be taking the pickup loop spaced $2z$ from the field coil. Taking $2z=3\mu m$ gives 523$\Phi_0$/A vs 511$\Phi_0$/A for the circular loop model, and 504$\Phi_0$/A vs 489$\Phi_0$/A for the incomplete circular loop plus leads model: This approximation leads to about a 7\% contribution to the total error in the mutual inductance.

\subsubsection{Analytical approximations}

\begin{figure}
\includegraphics[width=3.3in,trim=0 0 0 0]{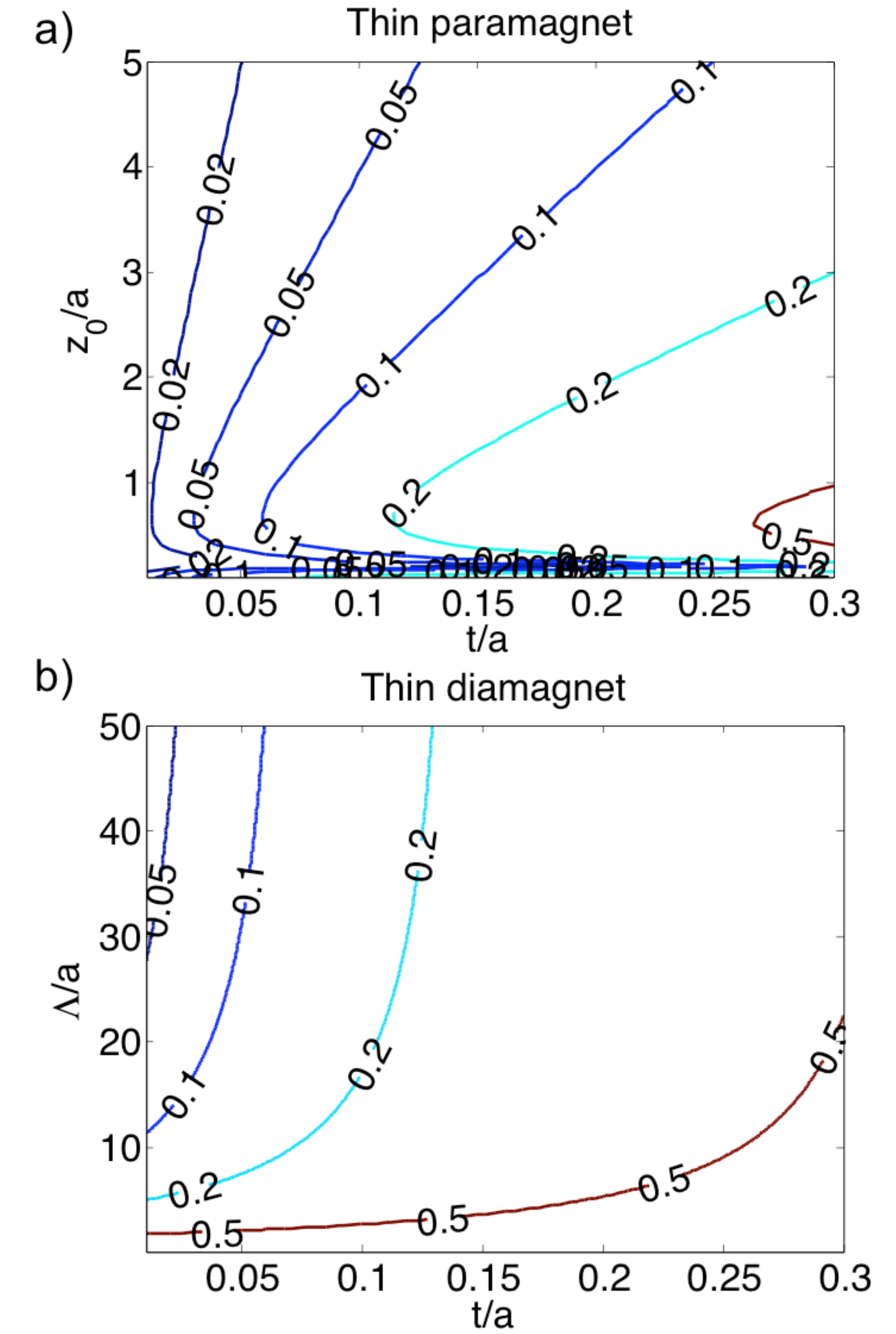}
\caption{(a) Fractional error $|\phi_{\rm exact}-\phi_{\rm analytical}|/\phi_{\rm exact}$ associated with using the thin paramagnetic (Table \ref{tab:analytical_limits}b) analytical expression instead of the exact expression Eq. \ref{eq:full_susc_expression}, assuming $\lambda\rightarrow\infty$ and $\mu_3=\mu_0$. (b) Fractional error for the thin diamagnetic (Table \ref{tab:analytical_limits}e) analytical expression, assuming $\mu_2=\mu_3=\mu_0$ and $z_0/a=0.2$.}

\label{fig:frac_error_1}
\end{figure}

\begin{figure}
\includegraphics[width=3.3in,trim=0 0 0 0]{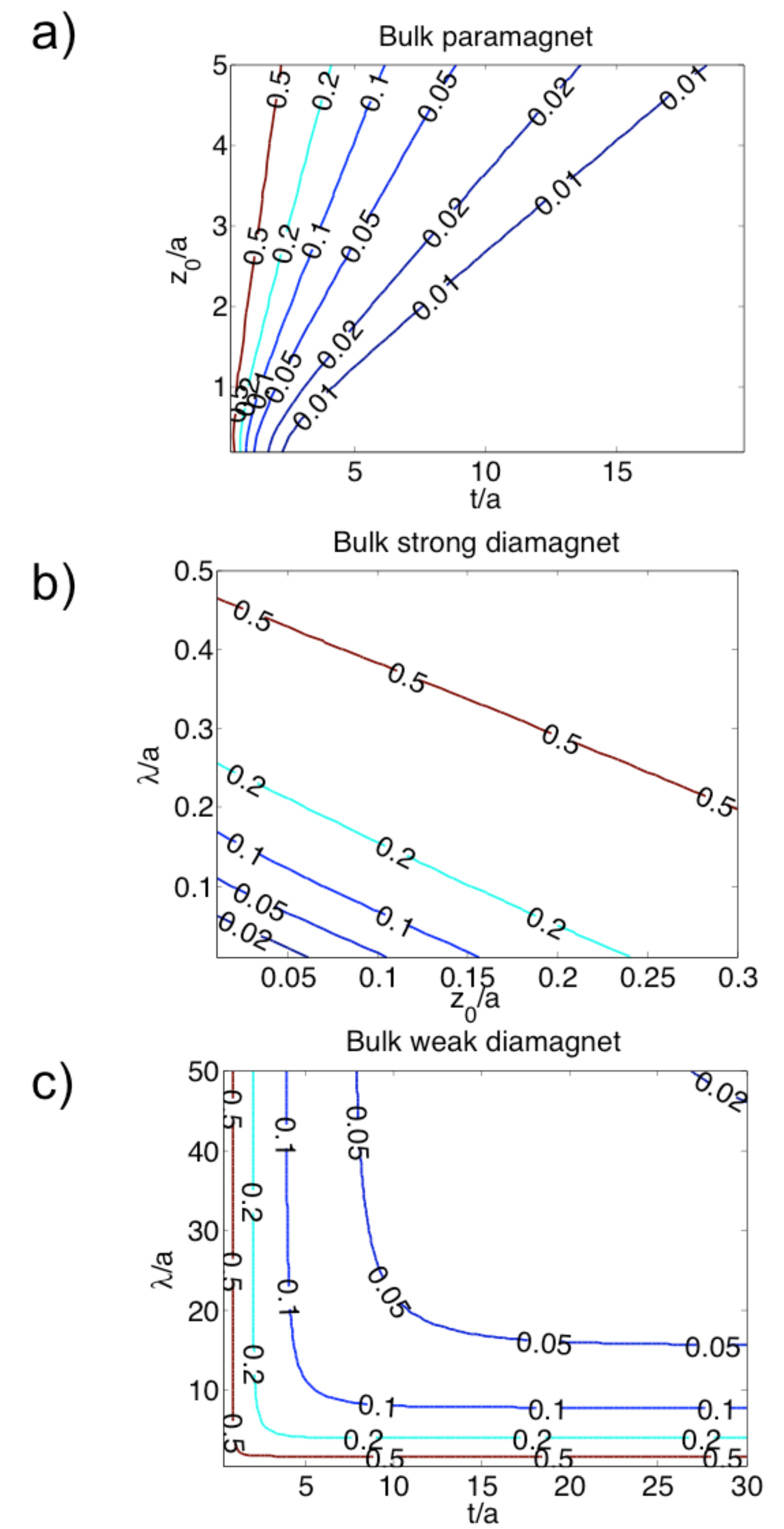}
\caption{ (a) Fractional error $|\phi_{\rm exact}-\phi_{\rm analytical}|/\phi_{\rm exact}$ associated with using the bulk paramagnetic analytic expression (Table \ref{tab:analytical_limits}a), assuming $\mu_3=\mu_0$ and $\lambda\rightarrow\infty$. (b) Fractional error for the bulk strong diamagnetic expression (Table \ref{tab:analytical_limits}c), assuming $t\rightarrow\infty$ and $\mu_2=\mu_3=\mu_0$. (c) Fractional error for the the bulk strong diamagnetic expression (Table \ref{tab:analytical_limits}d), assuming $z_0/a=0.2$, and $\mu_2=\mu_3=\mu_0$. }
\label{fig:frac_error_2}
\end{figure}

Table \ref{tab:analytical_limits} displays analytical approximations to the full expression Eq. \ref{eq:full_susc_expression} in various limits. These analytical approximations can be evaluated much faster than the full expression, and in some limits the temperature dependence of the penetration depth $\lambda$ or the permeabilty $\mu_2$ can be inferred from the data without curve fitting, aside from a multiplicative constant, but care must be taken. Figures \ref{fig:frac_error_1} and \ref{fig:frac_error_2} show contour plots of the fractional error $|\phi_{\rm exact}-\phi_{\rm analytical}|/\phi_{\rm exact}$ associated with using each of the 5 analytical expressions in Table \ref{tab:analytical_limits}. 

\label{sec:analytical_approx}

For the $\delta$-doped STO data of Fig. \ref{fig:introduction_figure4}a the best fit value for $\Lambda/a=113$, and $t/a\sim7\times10^{-4}$, so the error associated with using the analytical expression (Table \ref{tab:analytical_limits}e, Fig. \ref{fig:frac_error_1}b) is approximately 1\%. For the LaNiO$_3$ data of  Fig. \ref{fig:introduction_figure4}b $t/a=2.4\times10^{-3}$, and the error associated with using Table \ref{tab:analytical_limits}b, Fig. \ref{fig:frac_error_1}b is approximately 0.2\%. The systematic errors associated with using the analytical expressions for the 2-DEL data of Fig. \ref{fig:plot_susc_fits} are also negligible.

\subsubsection{Uncertainties in parameter values}
\label{sec:param_uncertainties}

\begin{figure}
\includegraphics[width=3in,trim=0 0 0 0]{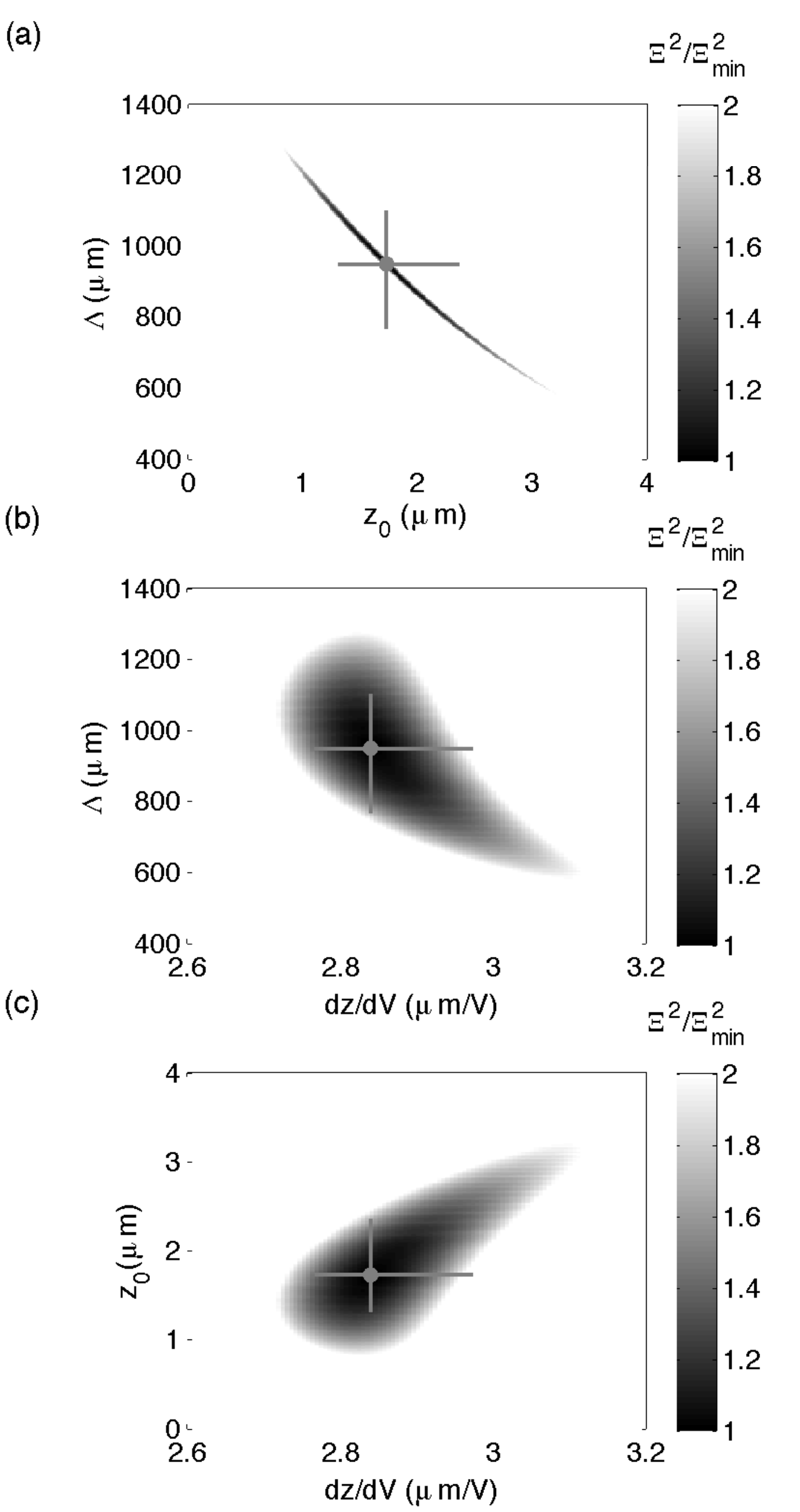}
\caption{Plot of the sum of the errors squared, divided by the global minimum  ($\Xi^2/\Xi^2_{\rm min}$), for the $\delta$-doped SrTiO$_3$ sample data of Figure \ref{fig:introduction_figure4}b, fit to the thin diamagnetic expression  (Table \ref{tab:analytical_limits}(e)), projecting the minimum value in the $\Lambda$, $z_0$, $dz/dV$ parameter space, taken along the third axis, onto the $\Lambda-z_0$ plane (a), the $\Lambda-dz/dV$ plane (b), and the $z_0-dz/dV$ plane (c), taking fixed values $a$=8.4$\mu$m and $b$=2.7$\mu$m. The global best fit values are $\Lambda=954\mu m$, $z_0=1.7\mu m$, and $dz/dV=2.8\mu m/V$. The solid symbols and lines are the best fit and 95\% confidence limits for the parameters from a statistical bootstrap analysis.}
\label{fig:plot_correlation_map_3d}
\end{figure}

The largest systematic errors in determining material parameters such as the penetration depth $\lambda$ and the permeability $\mu$ of a permeable superconductor are uncertainties in the parameters such as the height of the SQUID susceptometer $z_0$ above the sample surface, and the change in sensor height with applied voltage $dz/dV$. Figure \ref{fig:plot_correlation_map_3d} shows estimates for the uncertainties in the parameters $\Lambda$, $z_0$, and $dz/dV$ from fits to the $\delta$-doped STO data of Fig. \ref{fig:introduction_figure4}a. The gray-scale images in this figure display the error square sum $\Xi^2=\sum_n(\phi(n)-\phi_{\rm fit}(n))^2$ for a three dimensional volume in parameter space, projected onto the three 2-dimensional planes $\Lambda-z_0$, $z_0-dz/dV$, and $\Lambda-dz/dV$ by taking the minimum value of $\Xi^2$ along each projection axis. The other two parameters, a vertical shift $\delta\phi$ and a linear slope $\phi_{\rm linear} = \alpha z$, were optimized for each pixel in the 3-dimensional parameter space. One way to estimate the uncertainty in the parameters is to determine the region in parameter space where $\Xi^2$ is less than twice its minimum value. The global minimum value for $dz/dV$ (2.9$\mu$m/V) is consistent with our knowledge of the physical properties of our $z$-bender at low temperatures.  Fig. \ref{fig:plot_correlation_map_3d} shows that the best fit value for $\Lambda$ depends sensitively on $z_0$.  We estimate from our knowledge of the tip-sample geometry that the sensor height 1$\mu m < z_0 < 2.5\mu m$, which implies that 700$\mu m < \Lambda  < 1100\mu m$. As can be seen from Table \ref{tab:analytical_limits}e, the SQUID susceptibility in the thin diamagnetic limit is proportional to $a/\Lambda$, and therefore a systematic error in $a$ will result in a proportional error in $\Lambda$. We consider it unlikely that our estimate of $a$ is incorrect by more than $\pm$20\%, and therefore assign a further systematic error of $\pm$20\% to uncertainties in the effective sizes of the field coil and pickup loop. The Pearl length can be related to the density of superconducting carriers through $n_s=m^*/\mu_0e^2\Lambda$, where $e$ is the elementary charge. Using $m^*$=1.25$m_e$ \cite{kozuka2009two} results in $n_s = 3.8+2.5-1.1\times10^{12}$1/cm$^2$. 

\begin{figure}
\includegraphics[width=3in,trim=0 0 0 0]{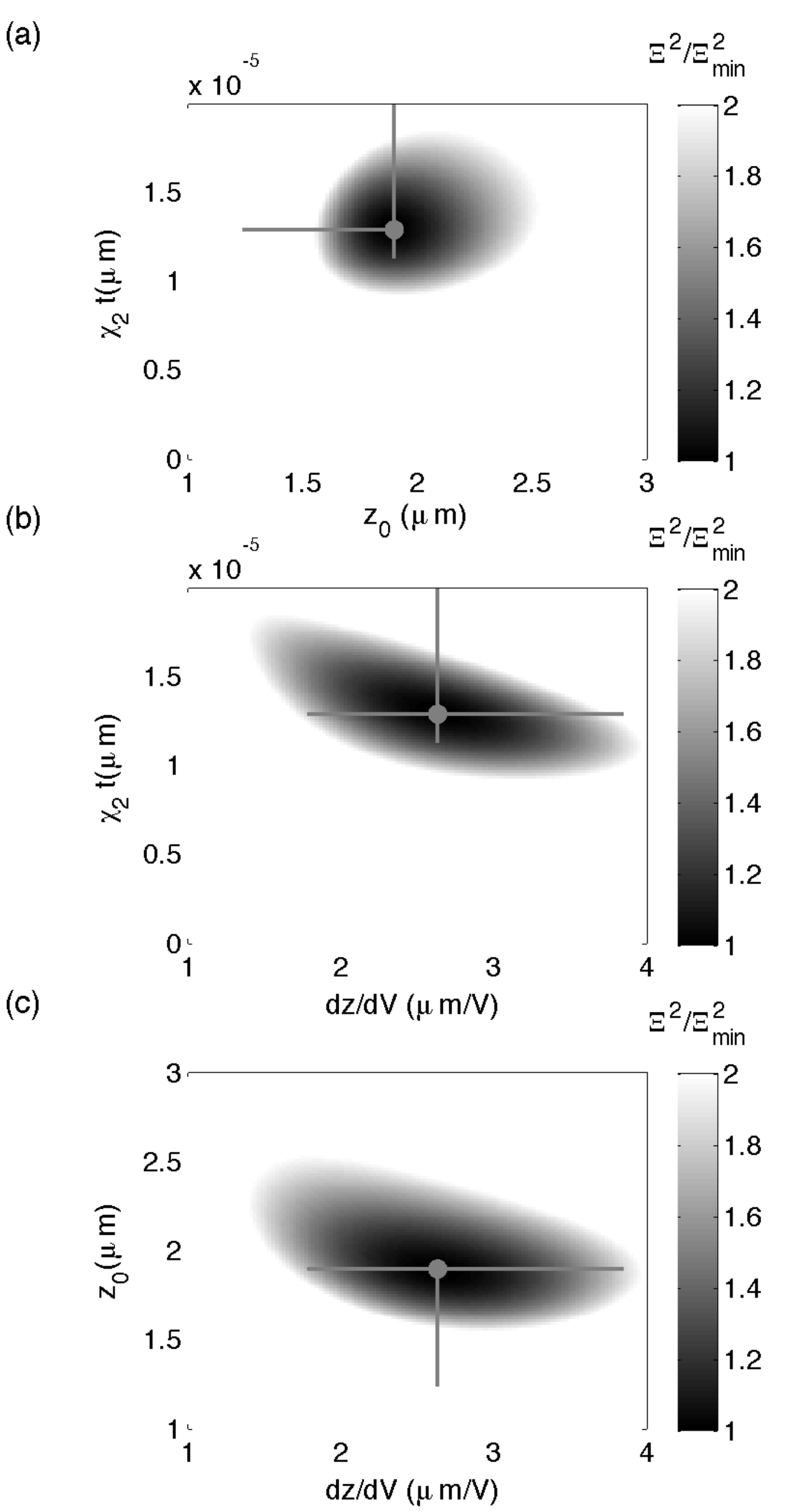}
\caption{Plot of the sum of squares error, divided by the global minimum  ($\Xi^2/\Xi^2_{\rm min}$), for fits of the LaNiO$_3$ data of Figure \ref{fig:introduction_figure4}d to the thin paramagnetic limit expression (Table \ref{tab:analytical_limits}(b)), projecting the minimum value in the $\chi_2t$, $z_0$, $dz/dV$ parameter space, taken along the third axis, onto the $\chi_2t-z_0$ plane (a), the $\chi_2t-dz/dV$ plane (b), and the $z_0-dz/dV$ plane (c), taking fixed values a=8.4$\mu$m and b=2.7$\mu$m. The global best fit values are $\chi_2t=1.3\times10^{-5}\mu$m, $z_0=1.9\mu$m, and $dz/dV=2.6\mu$m/V. The solid symbols and lines are most probable values and 95\% confidence limits for the parameters from a statistical bootstrap analysis.}
\label{fig:plot_correlation_map_3dmu}
\end{figure}

Fig. \ref{fig:plot_correlation_map_3dmu} displays the error square sum $\Xi^2$ for a three dimensional volume ($\chi_2t$, $z_0$, and $dz/dV$) in parameter space, projected onto the three 2-dimensional planes $\chi_2t-z_0$, $z_0-dz/dV$, and $\chi_2t-dz/dV$ for fits to the LaNiO$_3$ data of Fig. \ref{fig:introduction_figure4}b. If we assume that the susceptibility in LaNiO$_3$ arises from isolated paramagnetic spins, we can estimate the 2D substrate spin density $N_s$ by using $\chi_2 t=\mu_0 N_s (g \mu_B)^2 J(J+1)/3 k_B T$.\cite{bluhm2009spinlike} The systematic uncertainty in $\chi_2 t$ should again be proportional to our uncertainty in $a$. Assuming a $\pm$20\% uncertainty in $a$, $g=2$ and $J=1/2$ leads to $N_s \sim 6.4+5.1-2.3\times 10^{14}$ cm$^{-2}$: The diamagnetic signal in $\delta$-doped STO is 4000 times bigger than the paramagnetic signal in LaNiO$_3$, but the calculated superconducting carrier density is 70 times smaller than the calculated spin density. 

\subsection{Statistical errors}
\label{sec:statistics}
The solid symbols and lines overlaid on the gray scale images in Fig. \ref{fig:plot_correlation_map_3d} and \ref{fig:plot_correlation_map_3dmu} represent the best fit values and 95\% confidence intervals for the parameters using a statistical bootstrap analysis.\cite{efron1986bootstrap} Briefly, in this analysis a random sampling of the data was generated, with substitutions, to produce the same number of points as the original set. This set was fit to the model allowing all 5 parameters to vary, best fit parameters were recorded, and the procedure was repeated 5000 times. A histogram of the best fit parameters was generated, and confidence interval limits were set at the 2.5\% and 97.5\% levels. 

In the case of the $\delta$-doped STO data of Fig.'s \ref{fig:introduction_figure4}a and \ref{fig:plot_correlation_map_3d} it appears that the statistical uncertainties are smaller than the uncertainties associated with our imprecise knowledge of the sensor height $z_0$. For the case of LaNiO$_3$ of Fig.'s \ref{fig:introduction_figure4}b and \ref{fig:plot_correlation_map_3dmu} the bootstrap statistical analysis indicates that the statistical uncertainties dominate, as might be expected from the noise in the data.


\bibliographystyle{apsrev}
\bibliography{slab_susc}

\begin{thebibliography}{28}
\expandafter\ifx\csname natexlab\endcsname\relax\def\natexlab#1{#1}\fi
\expandafter\ifx\csname bibnamefont\endcsname\relax
  \def\bibnamefont#1{#1}\fi
\expandafter\ifx\csname bibfnamefont\endcsname\relax
  \def\bibfnamefont#1{#1}\fi
\expandafter\ifx\csname citenamefont\endcsname\relax
  \def\citenamefont#1{#1}\fi
\expandafter\ifx\csname url\endcsname\relax
  \def\url#1{\texttt{#1}}\fi
\expandafter\ifx\csname urlprefix\endcsname\relax\def\urlprefix{URL }\fi
\providecommand{\bibinfo}[2]{#2}
\providecommand{\eprint}[2][]{\url{#2}}

\bibitem[{\citenamefont{Kirtley and Wikswo}(1999)}]{kirtley1999ssm}
\bibinfo{author}{\bibfnamefont{J.~R.} \bibnamefont{Kirtley}} \bibnamefont{and}
  \bibinfo{author}{\bibfnamefont{J.}~\bibnamefont{Wikswo}},
  \bibinfo{journal}{Ann. Rev. Mat. Sci.} \textbf{\bibinfo{volume}{29}},
  \bibinfo{pages}{117} (\bibinfo{year}{1999}).

\bibitem[{\citenamefont{Kirtley}(2010)}]{kirtley2010fundamental}
\bibinfo{author}{\bibfnamefont{J.~R.} \bibnamefont{Kirtley}},
  \bibinfo{journal}{Reports on Progress in Physics}
  \textbf{\bibinfo{volume}{73}}, \bibinfo{pages}{126501}
  (\bibinfo{year}{2010}),
  \urlprefix\url{http://stacks.iop.org/0034-4885/73/i=12/a=126501}.

\bibitem[{\citenamefont{Gardner et~al.}(2001)\citenamefont{Gardner, Wynn,
  Bj{\"o}rnsson, Straver, Moler, Kirtley, and Ketchen}}]{gardner2001scanning}
\bibinfo{author}{\bibfnamefont{B.~W.} \bibnamefont{Gardner}},
  \bibinfo{author}{\bibfnamefont{J.~C.} \bibnamefont{Wynn}},
  \bibinfo{author}{\bibfnamefont{P.~G.} \bibnamefont{Bj{\"o}rnsson}},
  \bibinfo{author}{\bibfnamefont{E.~W.~J.} \bibnamefont{Straver}},
  \bibinfo{author}{\bibfnamefont{K.~A.} \bibnamefont{Moler}},
  \bibinfo{author}{\bibfnamefont{J.~R.} \bibnamefont{Kirtley}},
  \bibnamefont{and} \bibinfo{author}{\bibfnamefont{M.~B.}
  \bibnamefont{Ketchen}}, \bibinfo{journal}{Rev. Sci. Instr.}
  \textbf{\bibinfo{volume}{72}}, \bibinfo{pages}{2361} (\bibinfo{year}{2001}).

\bibitem[{\citenamefont{Tafuri et~al.}(2004)\citenamefont{Tafuri, Kirtley,
  Medaglia, Orgiani, and Balestrino}}]{tafuri2004magnetic}
\bibinfo{author}{\bibfnamefont{F.}~\bibnamefont{Tafuri}},
  \bibinfo{author}{\bibfnamefont{J.~R.} \bibnamefont{Kirtley}},
  \bibinfo{author}{\bibfnamefont{P.~G.} \bibnamefont{Medaglia}},
  \bibinfo{author}{\bibfnamefont{P.}~\bibnamefont{Orgiani}}, \bibnamefont{and}
  \bibinfo{author}{\bibfnamefont{G.}~\bibnamefont{Balestrino}},
  \bibinfo{journal}{Phys. Rev. Lett.} \textbf{\bibinfo{volume}{92}},
  \bibinfo{pages}{157006} (\bibinfo{year}{2004}).

\bibitem[{\citenamefont{Hicks et~al.}(2009)\citenamefont{Hicks, Lippman, Huber,
  Analytis, Chu, Erickson, Fisher, and Moler}}]{hicks2009evidence}
\bibinfo{author}{\bibfnamefont{C.~W.} \bibnamefont{Hicks}},
  \bibinfo{author}{\bibfnamefont{T.~M.} \bibnamefont{Lippman}},
  \bibinfo{author}{\bibfnamefont{M.~E.} \bibnamefont{Huber}},
  \bibinfo{author}{\bibfnamefont{J.~G.} \bibnamefont{Analytis}},
  \bibinfo{author}{\bibfnamefont{J.~H.} \bibnamefont{Chu}},
  \bibinfo{author}{\bibfnamefont{A.~S.} \bibnamefont{Erickson}},
  \bibinfo{author}{\bibfnamefont{I.~R.} \bibnamefont{Fisher}},
  \bibnamefont{and} \bibinfo{author}{\bibfnamefont{K.~A.} \bibnamefont{Moler}},
  \bibinfo{journal}{Phys. Rev. Lett.} \textbf{\bibinfo{volume}{103}},
  \bibinfo{pages}{127003} (\bibinfo{year}{2009}).

\bibitem[{\citenamefont{Luan et~al.}(2011)\citenamefont{Luan, Lippman, Hicks,
  Bert, Auslaender, Chu, Analytis, Fisher, and Moler}}]{luan2011local}
\bibinfo{author}{\bibfnamefont{L.}~\bibnamefont{Luan}},
  \bibinfo{author}{\bibfnamefont{T.~M.} \bibnamefont{Lippman}},
  \bibinfo{author}{\bibfnamefont{C.~W.} \bibnamefont{Hicks}},
  \bibinfo{author}{\bibfnamefont{J.~A.} \bibnamefont{Bert}},
  \bibinfo{author}{\bibfnamefont{O.~M.} \bibnamefont{Auslaender}},
  \bibinfo{author}{\bibfnamefont{J.~H.} \bibnamefont{Chu}},
  \bibinfo{author}{\bibfnamefont{J.~G.} \bibnamefont{Analytis}},
  \bibinfo{author}{\bibfnamefont{I.~R.} \bibnamefont{Fisher}},
  \bibnamefont{and} \bibinfo{author}{\bibfnamefont{K.~A.} \bibnamefont{Moler}},
  \bibinfo{journal}{Phys. Rev. Lett.} \textbf{\bibinfo{volume}{106}},
  \bibinfo{pages}{67001} (\bibinfo{year}{2011}).

\bibitem[{\citenamefont{Kalisky et~al.}(2010)\citenamefont{Kalisky, Kirtley,
  Analytis, Chu, Vailionis, Fisher, and Moler}}]{kalisky2010stripes}
\bibinfo{author}{\bibfnamefont{B.}~\bibnamefont{Kalisky}},
  \bibinfo{author}{\bibfnamefont{J.~R.} \bibnamefont{Kirtley}},
  \bibinfo{author}{\bibfnamefont{J.~G.} \bibnamefont{Analytis}},
  \bibinfo{author}{\bibfnamefont{J.~H.} \bibnamefont{Chu}},
  \bibinfo{author}{\bibfnamefont{A.}~\bibnamefont{Vailionis}},
  \bibinfo{author}{\bibfnamefont{I.~R.} \bibnamefont{Fisher}},
  \bibnamefont{and} \bibinfo{author}{\bibfnamefont{K.~A.} \bibnamefont{Moler}},
  \bibinfo{journal}{Phys. Rev. B} \textbf{\bibinfo{volume}{81}},
  \bibinfo{pages}{184513} (\bibinfo{year}{2010}).

\bibitem[{\citenamefont{Bluhm et~al.}(2009)\citenamefont{Bluhm, Bert, Koshnick,
  Huber, and Moler}}]{bluhm2009spinlike}
\bibinfo{author}{\bibfnamefont{H.}~\bibnamefont{Bluhm}},
  \bibinfo{author}{\bibfnamefont{J.~A.} \bibnamefont{Bert}},
  \bibinfo{author}{\bibfnamefont{N.~C.} \bibnamefont{Koshnick}},
  \bibinfo{author}{\bibfnamefont{M.~E.} \bibnamefont{Huber}}, \bibnamefont{and}
  \bibinfo{author}{\bibfnamefont{K.~A.} \bibnamefont{Moler}},
  \bibinfo{journal}{Phys. Rev. Lett.} \textbf{\bibinfo{volume}{103}},
  \bibinfo{pages}{26805} (\bibinfo{year}{2009}).

\bibitem[{\citenamefont{Bert et~al.}(2011)\citenamefont{Bert, Kalisky, Bell,
  Kim, Hikita, Hwang, and Moler}}]{bert2011direct}
\bibinfo{author}{\bibfnamefont{J.~A.} \bibnamefont{Bert}},
  \bibinfo{author}{\bibfnamefont{B.}~\bibnamefont{Kalisky}},
  \bibinfo{author}{\bibfnamefont{C.}~\bibnamefont{Bell}},
  \bibinfo{author}{\bibfnamefont{M.}~\bibnamefont{Kim}},
  \bibinfo{author}{\bibfnamefont{Y.}~\bibnamefont{Hikita}},
  \bibinfo{author}{\bibfnamefont{H.~Y.} \bibnamefont{Hwang}}, \bibnamefont{and}
  \bibinfo{author}{\bibfnamefont{K.~A.} \bibnamefont{Moler}},
  \bibinfo{journal}{Nature Physics}  (\bibinfo{year}{2011}).

\bibitem[{\citenamefont{Kalisky et~al.}(2011)\citenamefont{Kalisky, Kirtley,
  Analytis, Chu, Fisher, and Moler}}]{kalisky2011para}
\bibinfo{author}{\bibfnamefont{B.}~\bibnamefont{Kalisky}},
  \bibinfo{author}{\bibfnamefont{J.~R.} \bibnamefont{Kirtley}},
  \bibinfo{author}{\bibfnamefont{J.~G.} \bibnamefont{Analytis}},
  \bibinfo{author}{\bibfnamefont{J.~H.} \bibnamefont{Chu}},
  \bibinfo{author}{\bibfnamefont{I.~R.} \bibnamefont{Fisher}},
  \bibnamefont{and} \bibinfo{author}{\bibfnamefont{K.~A.} \bibnamefont{Moler}},
  \bibinfo{journal}{to be published}  (\bibinfo{year}{2011}).

\bibitem[{\citenamefont{Prozorov and Giannetta}(2006)}]{prozorov2006magnetic}
\bibinfo{author}{\bibfnamefont{R.}~\bibnamefont{Prozorov}} \bibnamefont{and}
  \bibinfo{author}{\bibfnamefont{R.~W.} \bibnamefont{Giannetta}},
  \bibinfo{journal}{Supercond. Sci. Tech.} \textbf{\bibinfo{volume}{19}},
  \bibinfo{pages}{R41} (\bibinfo{year}{2006}).

\bibitem[{\citenamefont{Prozorov et~al.}(2000)\citenamefont{Prozorov,
  Giannetta, Fournier, and Greene}}]{prozorov2000evidence}
\bibinfo{author}{\bibfnamefont{R.}~\bibnamefont{Prozorov}},
  \bibinfo{author}{\bibfnamefont{R.~W.} \bibnamefont{Giannetta}},
  \bibinfo{author}{\bibfnamefont{P.}~\bibnamefont{Fournier}}, \bibnamefont{and}
  \bibinfo{author}{\bibfnamefont{R.~L.} \bibnamefont{Greene}},
  \bibinfo{journal}{Phys. Rev. Lett.} \textbf{\bibinfo{volume}{85}},
  \bibinfo{pages}{3700} (\bibinfo{year}{2000}),
  \urlprefix\url{http://link.aps.org/doi/10.1103/PhysRevLett.85.3700}.

\bibitem[{\citenamefont{Kogan}(2003)}]{kogan2003meissner}
\bibinfo{author}{\bibfnamefont{V.~G.} \bibnamefont{Kogan}},
  \bibinfo{journal}{Phys. Rev. B} \textbf{\bibinfo{volume}{68}},
  \bibinfo{pages}{104511} (\bibinfo{year}{2003}).

\bibitem[{\citenamefont{Fiory et~al.}(1988)\citenamefont{Fiory, Hebard,
  Mankiewich, and Howard}}]{fiory1988penetration}
\bibinfo{author}{\bibfnamefont{A.}~\bibnamefont{Fiory}},
  \bibinfo{author}{\bibfnamefont{A.}~\bibnamefont{Hebard}},
  \bibinfo{author}{\bibfnamefont{P.}~\bibnamefont{Mankiewich}},
  \bibnamefont{and} \bibinfo{author}{\bibfnamefont{R.}~\bibnamefont{Howard}},
  \bibinfo{journal}{Applied physics letters} \textbf{\bibinfo{volume}{52}},
  \bibinfo{pages}{2165} (\bibinfo{year}{1988}).

\bibitem[{\citenamefont{Lee et~al.}(1994)\citenamefont{Lee, Paget, Lemberger,
  Foltyn, and Wu}}]{lee1994crossover}
\bibinfo{author}{\bibfnamefont{J.}~\bibnamefont{Lee}},
  \bibinfo{author}{\bibfnamefont{K.}~\bibnamefont{Paget}},
  \bibinfo{author}{\bibfnamefont{T.}~\bibnamefont{Lemberger}},
  \bibinfo{author}{\bibfnamefont{S.}~\bibnamefont{Foltyn}}, \bibnamefont{and}
  \bibinfo{author}{\bibfnamefont{X.}~\bibnamefont{Wu}},
  \bibinfo{journal}{Physical Review B} \textbf{\bibinfo{volume}{50}},
  \bibinfo{pages}{3337} (\bibinfo{year}{1994}).

\bibitem[{\citenamefont{Claassen et~al.}(1997)\citenamefont{Claassen, Wilson,
  Byers, and Adrian}}]{claassen1997optimizing}
\bibinfo{author}{\bibfnamefont{J.}~\bibnamefont{Claassen}},
  \bibinfo{author}{\bibfnamefont{M.}~\bibnamefont{Wilson}},
  \bibinfo{author}{\bibfnamefont{J.}~\bibnamefont{Byers}}, \bibnamefont{and}
  \bibinfo{author}{\bibfnamefont{S.}~\bibnamefont{Adrian}},
  \bibinfo{journal}{Journal of applied physics} \textbf{\bibinfo{volume}{82}},
  \bibinfo{pages}{3028} (\bibinfo{year}{1997}).

\bibitem[{\citenamefont{Ohtomo and Hwang}(2004)}]{ohtomo2004high}
\bibinfo{author}{\bibfnamefont{A.}~\bibnamefont{Ohtomo}} \bibnamefont{and}
  \bibinfo{author}{\bibfnamefont{H.~Y.} \bibnamefont{Hwang}},
  \bibinfo{journal}{Nature} \textbf{\bibinfo{volume}{427}},
  \bibinfo{pages}{423} (\bibinfo{year}{2004}).

\bibitem[{\citenamefont{Huber et~al.}(2008)\citenamefont{Huber, Koshnick,
  Bluhm, Archuleta, Azua, Bj{\"o}rnsson, Gardner, Halloran, Lucero, and
  Moler}}]{huber2008gradiometric}
\bibinfo{author}{\bibfnamefont{M.~E.} \bibnamefont{Huber}},
  \bibinfo{author}{\bibfnamefont{N.~C.} \bibnamefont{Koshnick}},
  \bibinfo{author}{\bibfnamefont{H.}~\bibnamefont{Bluhm}},
  \bibinfo{author}{\bibfnamefont{L.~J.} \bibnamefont{Archuleta}},
  \bibinfo{author}{\bibfnamefont{T.}~\bibnamefont{Azua}},
  \bibinfo{author}{\bibfnamefont{P.~G.} \bibnamefont{Bj{\"o}rnsson}},
  \bibinfo{author}{\bibfnamefont{B.~W.} \bibnamefont{Gardner}},
  \bibinfo{author}{\bibfnamefont{S.~T.} \bibnamefont{Halloran}},
  \bibinfo{author}{\bibfnamefont{E.~A.} \bibnamefont{Lucero}},
  \bibnamefont{and} \bibinfo{author}{\bibfnamefont{K.~A.} \bibnamefont{Moler}},
  \bibinfo{journal}{Rev. Sci. Instr.} \textbf{\bibinfo{volume}{79}},
  \bibinfo{pages}{053704} (\bibinfo{year}{2008}).

\bibitem[{\citenamefont{Ketchen and Kirtley}(1995)}]{ketchen1995design}
\bibinfo{author}{\bibfnamefont{M.~B.} \bibnamefont{Ketchen}} \bibnamefont{and}
  \bibinfo{author}{\bibfnamefont{J.~R.} \bibnamefont{Kirtley}},
  \bibinfo{journal}{IEEE Trans. Appl. Supercond.} \textbf{\bibinfo{volume}{5}},
  \bibinfo{pages}{2133} (\bibinfo{year}{1995}).

\bibitem[{\citenamefont{Bluhm}(2007)}]{bluhm2007magnetic}
\bibinfo{author}{\bibfnamefont{H.}~\bibnamefont{Bluhm}},
  \bibinfo{journal}{Phys. Rev. B} \textbf{\bibinfo{volume}{76}},
  \bibinfo{pages}{144507} (\bibinfo{year}{2007}).

\bibitem[{\citenamefont{Matsumoto et~al.}(1982)\citenamefont{Matsumoto,
  Umezawa, and Tachiki}}]{matsumoto1982parameters}
\bibinfo{author}{\bibfnamefont{H.}~\bibnamefont{Matsumoto}},
  \bibinfo{author}{\bibfnamefont{H.}~\bibnamefont{Umezawa}}, \bibnamefont{and}
  \bibinfo{author}{\bibfnamefont{M.}~\bibnamefont{Tachiki}},
  \bibinfo{journal}{Phys. Rev. B} \textbf{\bibinfo{volume}{25}},
  \bibinfo{pages}{6633} (\bibinfo{year}{1982}),
  \urlprefix\url{http://link.aps.org/doi/10.1103/PhysRevB.25.6633}.

\bibitem[{\citenamefont{Gray}(1983)}]{gray1983ginzburg}
\bibinfo{author}{\bibfnamefont{K.~E.} \bibnamefont{Gray}},
  \bibinfo{journal}{Phys. Rev. B} \textbf{\bibinfo{volume}{27}},
  \bibinfo{pages}{4157} (\bibinfo{year}{1983}),
  \urlprefix\url{http://link.aps.org/doi/10.1103/PhysRevB.27.4157}.

\bibitem[{\citenamefont{Buzdin and
  Bulaevski{\u\i}}(1986)}]{buzdin1986antiferromagnetic}
\bibinfo{author}{\bibfnamefont{A.}~\bibnamefont{Buzdin}} \bibnamefont{and}
  \bibinfo{author}{\bibfnamefont{L.}~\bibnamefont{Bulaevski{\u\i}}},
  \bibinfo{journal}{Soviet Physics Uspekhi} \textbf{\bibinfo{volume}{29}},
  \bibinfo{pages}{412} (\bibinfo{year}{1986}).

\bibitem[{\citenamefont{Bj{\"o}rnsson et~al.}(2001)\citenamefont{Bj{\"o}rnsson,
  Gardner, Kirtley, and Moler}}]{bjornsson2001scanning}
\bibinfo{author}{\bibfnamefont{P.~G.} \bibnamefont{Bj{\"o}rnsson}},
  \bibinfo{author}{\bibfnamefont{B.~W.} \bibnamefont{Gardner}},
  \bibinfo{author}{\bibfnamefont{J.~R.} \bibnamefont{Kirtley}},
  \bibnamefont{and} \bibinfo{author}{\bibfnamefont{K.~A.} \bibnamefont{Moler}},
  \bibinfo{journal}{Rev. of Sci. Instr.} \textbf{\bibinfo{volume}{72}},
  \bibinfo{pages}{4153} (\bibinfo{year}{2001}).

\bibitem[{\citenamefont{Kozuka et~al.}(2009)\citenamefont{Kozuka, Kim, Bell,
  Kim, Hikita, and Hwang}}]{kozuka2009two}
\bibinfo{author}{\bibfnamefont{Y.}~\bibnamefont{Kozuka}},
  \bibinfo{author}{\bibfnamefont{M.}~\bibnamefont{Kim}},
  \bibinfo{author}{\bibfnamefont{C.}~\bibnamefont{Bell}},
  \bibinfo{author}{\bibfnamefont{B.}~\bibnamefont{Kim}},
  \bibinfo{author}{\bibfnamefont{Y.}~\bibnamefont{Hikita}}, \bibnamefont{and}
  \bibinfo{author}{\bibfnamefont{H.}~\bibnamefont{Hwang}},
  \bibinfo{journal}{Nature} \textbf{\bibinfo{volume}{462}},
  \bibinfo{pages}{487} (\bibinfo{year}{2009}).

\bibitem[{\citenamefont{Brandt and Clem}(2004)}]{brandt2004superconducting}
\bibinfo{author}{\bibfnamefont{E.~H.} \bibnamefont{Brandt}} \bibnamefont{and}
  \bibinfo{author}{\bibfnamefont{J.~R.} \bibnamefont{Clem}},
  \bibinfo{journal}{Phys. Rev. B} \textbf{\bibinfo{volume}{69}},
  \bibinfo{pages}{184509} (\bibinfo{year}{2004}),
  \urlprefix\url{http://link.aps.org/doi/10.1103/PhysRevB.69.184509}.

\bibitem[{\citenamefont{Caviglia et~al.}(2010)\citenamefont{Caviglia, Gariglio,
  Cancellieri, Sac{\'e}p{\'e}, F{\^e}te, Reyren, Gabay, Morpurgo, and
  Triscone}}]{caviglia2010two}
\bibinfo{author}{\bibfnamefont{A.~D.} \bibnamefont{Caviglia}},
  \bibinfo{author}{\bibfnamefont{S.}~\bibnamefont{Gariglio}},
  \bibinfo{author}{\bibfnamefont{C.}~\bibnamefont{Cancellieri}},
  \bibinfo{author}{\bibfnamefont{B.}~\bibnamefont{Sac{\'e}p{\'e}}},
  \bibinfo{author}{\bibfnamefont{A.}~\bibnamefont{F{\^e}te}},
  \bibinfo{author}{\bibfnamefont{N.}~\bibnamefont{Reyren}},
  \bibinfo{author}{\bibfnamefont{M.}~\bibnamefont{Gabay}},
  \bibinfo{author}{\bibfnamefont{A.~F.} \bibnamefont{Morpurgo}},
  \bibnamefont{and} \bibinfo{author}{\bibfnamefont{J.~M.}
  \bibnamefont{Triscone}}, \bibinfo{journal}{Phys. Rev. Lett.}
  \textbf{\bibinfo{volume}{105}}, \bibinfo{pages}{236802}
  (\bibinfo{year}{2010}).

\bibitem[{\citenamefont{Efron and Tibshirani}(1986)}]{efron1986bootstrap}
\bibinfo{author}{\bibfnamefont{B.}~\bibnamefont{Efron}} \bibnamefont{and}
  \bibinfo{author}{\bibfnamefont{R.}~\bibnamefont{Tibshirani}},
  \bibinfo{journal}{Statistical science} \textbf{\bibinfo{volume}{1}},
  \bibinfo{pages}{54} (\bibinfo{year}{1986}).

\end{thebibliography}

\end{document}